\documentclass[aps,pra,reprint,groupedaddress]{revtex4-2}

\usepackage[T1]{fontenc}
\usepackage{graphicx}
\usepackage{amsmath}
\usepackage{amssymb}
\usepackage{braket}
\usepackage{dsfont}
\usepackage{algpseudocode}

\newcounter{algorithm}
\renewcommand{\thealgorithm}{\arabic{algorithm}}

\makeatletter
\newcommand{\fnum@algorithm}{Algorithm~\thealgorithm}

\newcommand{\ext@algorithm}{loa}

\newenvironment{algorithm}[1][tbp]{%
    \begin{figure}[#1]%
    \def\@captype{algorithm}%
    \par
    \noindent\rule{\linewidth}{0.4pt}%
    \par\vspace{0.5ex}%
}{%
    \par\vspace{0.5ex}%
    \noindent\rule{\linewidth}{0.4pt}%
    \par
    \end{figure}%
}

\makeatother

\begin{document}
\title{Optimal Interpolation of Entanglement Purification Protocols}

\author{Matthew Barber}
\affiliation{Department of Computer Science, University of York, York YO10 5GH, United Kingdom}

\author{Stefano Pirandola}
\affiliation{Department of Computer Science, University of York, York YO10 5GH, United Kingdom}

\begin{abstract}
    Bipartite entanglement purification is the conversion of copies of weakly entangled pairs shared between two separated parties into a smaller number of strongly entangled shared pairs using only local operations and classical communication.
    Choosing between different entanglement purification protocols generally involves weighing up a trade-off between the ratio of strongly entangled pairs produced to weakly entangled pairs consumed, which we call the rate of the protocol, and the degree of the entanglement of the strongly entangled pairs, typically measured by the fidelity of those pairs to maximally entangled states.
    By randomly choosing a protocol according to a probability distribution over a list of protocols for each pair we want to produce, we can achieve rates and fidelities not achieved by any of the original protocols.
    Here, we show how to choose this distribution to maximize the rate at which we produce qubit pairs with a given fidelity to a Bell state or, equivalently, to maximize the fidelity to a Bell state of the qubit pairs produced at a given rate. We investigate both the asymptotic case, where the number of initial pairs goes to infinity, and the finite-size regime, where protocols are restricted to a finite number of weakly entangled pairs.
\end{abstract}

\maketitle

\section{Introduction}

In quantum information, we have resources with no counterpart in classical information.
Of these, entanglement is one of the better understood.
Moreover, it is one of the most useful;
among other applications, entanglement is used in protocols to teleport quantum states~\cite{QuantumTeleportation}, double the amount of classical information that can be communicated per use of a noiseless qubit channel \cite{SuperDenseCoding} and share secret cryptographic keys~\cite{BB84,Ekert1991,pirandola2020advances}. 
Like other quantum resources, it is limited by a fundamental rate-loss scaling law~\cite{pirandola2017fundamental}, thus requiring the use of quantum networks for long-distance end-to-end distribution.
\par
Entanglement can only be created locally \cite{EntanglementReview}. Therefore, two separated parties, Alice and Bob, typically come to share entanglement by an entangled pair being created locally and then the two halves being sent to Alice and Bob.
This process introduces noise in the entangled pair and so we cannot expect Alice and Bob's state to be maximally entangled.
However, if this process produces pairs whose entanglement is of lower quality than we require or produces pairs at a higher rate than we need to use them, Alice and Bob can choose to sacrifice some of their entangled pairs in the hope of increasing the quality of the entanglement of the others.
This process is known as entanglement distillation or entanglement purification~\cite{EntanglementOfFormation}.
\par
We will not be concerned here with the quantum Shannon-theoretic task of converting our noisy entangled pairs into as many perfect maximally entangled states as possible in the limit as the number of pairs we have to work with goes to infinity.
Rather, we will be focused on those protocols which achieve some non-maximal level of entanglement, measured by the fidelity of the resulting states with a maximally entangled pair.
Moreover, we will focus specifically on the typical case where our initial states are independent qubit pairs.
\par
Generally, there will be a trade-off between the number of pairs we expect to purify per initial pair consumed and the fidelity of the resulting pairs with the chosen Bell state, known as the Bell fidelity, and different purification protocols will provide us with different trade-off options.
For some applications, particularly in the context of quantum networks, we may want to convert our initial states into purified states at least at a particular rate or produce purified states with at least a particular Bell fidelity. 
For instance, our application might require a minimum quality of entanglement~\cite{ContinuousDistribution} or there may be a bottleneck elsewhere in the network that means Alice and Bob can, up to a point, lower the rate at which they purify pairs with little to no downside~\cite{RateBottleneck}.
\par
By randomly choosing a purification protocol for each pair we produce, which we shall refer to as interpolating between protocols, we can achieve trade-offs in between the options provided by our original protocols.
A preliminary approach to protocol interpolation was discussed in Ref.~\cite{DistillationInterpolation}.
However, in that paper, interpolation was only considered between protocols that all aimed to convert the same, fixed number of initial pairs into one purified pair.
Here, we consider an optimal strategy for interpolation between protocols that may consume a variable number of pairs to produce one purified pair.
\par
If we have a list of entanglement purification protocols and want to interpolate between them, we must choose a probability distribution over them which tells us with what probability we will choose any given protocol from our list to purify our next pair.
Different options for this probability distribution result in different trade-offs between the rate of and fidelity achieved by the corresponding ``interpolated'' protocol.
In Section~\ref{AsymptoticInterpolation}, we show how to find the probability distribution which produces pairs with the highest Bell fidelity at a particular rate or produces pairs with a particular Bell fidelity at the highest rate. This is done in the limit as the number of initial pairs we have to purify goes to infinity.
\par
Then, in Section~\ref{Finite-Size}, we investigate the performance of interpolated purification protocols in the case where we only have a fixed, finite number of pairs to purify.
Note that, although for convenience we will picture these initial pairs as all existing simultaneously in a ``pool'' of states to be purified, this analysis would also apply in other contexts where many of the pairs did not exist at the same time, for instance if two parties were frequently generating weakly entangled pairs, purifying them and then using the purified pairs for their purposes.
In that case, the finite number of initial pairs consumed would be the number of weakly entangled pairs generated in total, even though they need not all exist simultaneously.
\par
For concreteness, in both Sections \ref{AsymptoticInterpolation} and \ref{Finite-Size}, we demonstrate the application of the general techniques we develop to the case of interpolating between different numbers of iterations of the ``DEJMPS'' protocol \cite{DEJMPS} acting on initial pairs produced by sending halves of Bell states through simple noisy quantum channels.
That is, we analyse the performance of ``interpolated DEJMPS'' protocols acting on initial states of the form $\rho=\left(\mathds{1}_{A} \otimes \mathcal{E}_{B}\right)\left(\Ket{\Phi}\Bra{\Phi}\right)$, where $\Ket{\Phi}$ is our chosen Bell state, $\mathds{1}_{A}$ is the identity on the half of $\Ket{\Phi}$ that Alice keeps and $\mathcal{E}_{B}$ is a quantum channel applied to the second, transmitted qubit.
We compare the rates achieved by the interpolated protocols to the rates that would be achieved if we applied the DEJMPS protocol just enough times to reach a target fidelity of $0.9$ to $\Ket{\Phi}\Bra{\Phi}$.

\section{The Asymptotic Scenario}\label{AsymptoticInterpolation}

\subsection{The General Case}

\subsubsection{Achievable Rates and Fidelities} Suppose Alice and Bob have a pool of independent qubit pairs, each in some entangled state $\rho$, and want to convert them, via LOCC, into as many independent pairs as possible with as high a Bell fidelity as possible, where by the Bell fidelity we mean the fidelity to $\Ket{\Phi}\Bra{\Phi}$.
Naturally, we expect there to be a trade-off between the Bell fidelity they achieve and the number of such pairs they produce per pair consumed from the pool.
Here, we will define the fidelity between two density operators, $\tau$ and $\tau'$, to be
\begin{equation}
    F\left(\tau, \tau'\right) = \left(\mathrm{Tr}{\sqrt{\sqrt{\tau}\tau'\sqrt{\tau}}}\right)^{2}\text{.}
\end{equation}
Moreover, for every $F' \in \left[0, 1\right]$, we will let
\begin{equation}
    W_{F'} = \frac{4F'-1}{3}\Ket{\Phi}\Bra{\Phi} + \frac{4 - 4F'}{12}\mathds{1}_{4}\text{,}
\end{equation}
where $\mathds{1}_{4}$ is the $4$ by $4$ identity operator, denote the Werner state of Bell fidelity $F'$.
Now, let $H = \mathbb{C}^{2} \otimes \mathbb{C}^{2}$ be the Hilbert space in which the state of a qubit pair exists.
Then, given $F' \in \left(\frac{1}{2}, 1\right)$ and $R \in \left(0, \infty\right)$, if, for every $\delta \in {\left(0, R\right)}$ and $\epsilon \in \left(0, 1\right)$, there exists an $N \in \mathbb{N}$ and an LOCC, $\Lambda$, from the space of density operators on $H^{N}$ to the space of density operators on $H^{\lceil N\left(R-\delta\right) \rceil}$ such that \begin{equation}
F\left(\Lambda\left(\rho^{\otimes N}\right), W_{F'}^{\otimes\lceil N\left(R-\delta\right)\rceil}\right) \geq 1 - \epsilon\text{,}
\end{equation}
where $W_{F'}$ denotes the Werner state \cite{WernerState} of Bell fidelity $F'$, then we will say we can achieve a fidelity $F'$ with rate $R$ from $\rho$.

\subsubsection{The Interpolated Protocol}\label{InterpolationDefinitions} If Alice and Bob agree on a protocol which will, on average, consume some finite, non-zero number, $I$, of states from the pool and with probability $1$ will produce a single entangled qubit pair in some state $\rho'$, then we will say that the rate of this protocol is $\frac{1}{I}$.
By the law of large numbers, we can see that if Alice and Bob repeat the process enough times, they can ensure that the probability of the number of $\rho'$ states they produce divided by the number of $\rho$ states they consume being arbitrarily close to the rate of the protocol is arbitrarily close to $1$.
Now, imagine that we have a set of $K$ protocols, for some $K \in \mathbb{N}$, which produce a single copy of the states $\rho'_{1}, \rho'_{2}, \dots, \rho'_{K}$ respectively.
We will denote by $I_{1}, I_{2}, \dots, I_{K}$ the numbers of states the respective protocols consume on average and by $R_{1}, R_{2}, \dots, R_{K}$ the rates of our respective protocols.
\par
Let $p_{1}, p_{2}, \dots, p_{K}$ be the probabilities of a probability distribution over $\left\{1, 2, \dots, K\right\}$ and let
\begin{equation}
R = \frac{1}{p_{1}I_{1} + p_{2}I_{2} + \dots + p_{K}I_{K}}\text{.}\label{RateCalculation} 
\end{equation}
Now, choose $\delta \in \left(0, R\right)$ and $\epsilon \in {\left(0, 1\right)}$.
Given $N\ \in \mathbb{N}$, let $X_{1}, X_{2}, X_{3}, \dots, X_{\lceil N\left(R - \delta \right)\rceil}$ be independent random variables, each distributed over $\left\{1, 2, \dots, K\right\}$ according to the chosen probability distribution.
Now, imagine that Alice and Bob apply their $X_{1}$th protocol, followed by their $X_{2}$th protocol and so on until they apply the $X_{\lceil N\left(R - \delta \right)\rceil}$th protocol.
Let $Y_{\lceil N\left(R - \delta \right)\rceil}=0$ if the number of $\rho$ states consumed exceeds $N$ and let $Y_{\lceil N\left(R - \delta \right)\rceil}=1$ otherwise.
If $Y_{\lceil N\left(R - \delta \right)\rceil}=0$, then they declare failure and replace all of their qubits with $W_{\frac{1}{2}}$ or some other separable state of their choosing.
Otherwise, they declare success and apply the twirling operation, the operation described as a random bilateral rotation in \cite{BBPSSW}, to each of their pairs, converting them into Werner states while preserving each pair's Bell fidelity.
By the weak law of large numbers, as $N \to \infty$, $\Pr\left(Y_{\lceil N\left(R - \delta \right)\rceil}=1\right) \to 1$.
Therefore, we can fix $N$ so that $\Pr\left(Y_{\lceil N\left(R - \delta \right)\rceil}=1\right) \geq \sqrt{1-\epsilon}$.
\par
Let $F_{1}, \dots, F_{K}$ denote the Bell fidelities of $\rho'_{1}, \dots, \rho'_{K}$ respectively and let
\begin{equation}
F' = p_{1}F_{1} + p_{2}F_{2} + \dots p_{K}F_{K}\text{.}\label{FidelityCalculation}
\end{equation}
We then see that Alice and Bob's interpolated protocol maps $\rho^{\otimes N}$ to
\begin{equation}
\sigma^{M} = \Pr\left(Y_{M}=1\right)\tilde{W}_{F'}^{M} + \Pr\left(Y_{M}=0\right)W_{\frac{1}{2}}^{\otimes M}\text{,}
\end{equation}
where 
\begin{equation}
\tilde{W}_{F'}^{M} = \sum_{x \in \left\{1, \dots, K\right\}^{M}}\Pr\left(X^{M} = x \middle\vert Y_{M} = 1\right) \bigotimes_{i=1} ^{M}{W_{F_{x_{i}}}}\text{,}
\end{equation}
with $M = \lceil N\left(R-\delta\right) \rceil$ and $X^{M} = \left(X_{1}, \dots, X_{M}\right)$.
Note that both $\sigma^{M}$ and $W_{F}^{M}$ are Bell-diagonal.
\par
Let $Z_{1}, Z_{2}, \dots, Z_{M}$ be random variables on $\left\{1, 2, 3, 4\right\}$, conditionally independent given $\left(X_{1}, \dots, X_{M}\right)$, where, for every $i \in \left\{1, 2, \dots, M\right\}$ and $x \in \left\{1, 2, \dots, K\right\}^{M}$, \begin{equation}
\Pr\left(Z_{i} = 1 \middle \vert X^{M} = x\right) = F_{x_{i}}
\end{equation}
and
\begin{align}
\Pr\left(Z_{i} = 2 \middle \vert X^{M} = x\right) &= \Pr\left(Z_{i} = 3 \middle \vert X^{M} = x\right) \nonumber \\
&= \Pr\left(Z_{i} = 4 \middle \vert X^{M} = x\right) \nonumber \\
&= \frac{1-F_{x_{i}}}{3}\text{.}
\end{align}
We then associate with each possible sequence of $M$ Bell states a unique $z$ in $\left\{1, 2, 3, 4\right\}^{M}$ so that the probability mass function of $Z^{M}$ over $\left\{1, 2, 3, 4\right\}^{M}$ corresponds to diagonal elements of $W_{F'}^{M}$ in the Bell basis.
Likewise, we define a random variable, $Z'^{M}$, which is equal to $Z^{M}$ if $Y_{M}$ equals $1$ and, if $Y_{M}$ equals $0$, is a sequence of $M$ independent random variables over $\left\{1, 2, 3, 4\right\}$ which are each $1$ with probability $\frac{1}{2}$, $2$ with probability $\frac{1}{6}$, $3$ with probability $\frac{1}{6}$ and $4$ with probability $\frac{1}{6}$.
Then, the probability mass function of $Z'^{M}$ corresponds to the diagonal elements of $\sigma^{M}$ in the Bell basis just as the probability mass function of $Z^{M}$ did for $W_{F}^{M}$.
\par
Now,
\begin{align}
    &F\left(\sigma^{\lceil N \left(R-\delta\right) \rceil}, W_{F'}^{\otimes \lceil N \left(R-\delta\right) \rceil}\right) \nonumber \\
    = &\left(\sum_{z \in \left\{1, \dots, 4\right\}^{M}}{\sqrt{\Pr\left(Z^{M} = z\right) \Pr\left(Z'^{M} = z\right)}}\right)^{2} \nonumber \\
    \geq &\left(\sum_{z \in \left\{1, \dots, 4\right\}^{M}}{\sqrt{\Pr\left(Z^{M} = z\right) \Pr\left(Z^{M} = z, Y_{M}=1\right)}}\right)^{2} \nonumber \\
    \geq &\left(\sum_{z \in \left\{1, \dots, 4\right\}^{M}}{\Pr\left(Z^{M} = z, Y_{M}=1\right)}\right)^{2} \nonumber \\
    = &\Pr\left(Y_{M}=1\right)^{2} \nonumber \\
    \geq &1-\epsilon\text{.}\label{FidelityLowerBound}
\end{align}
Therefore, we can achieve a fidelity $F'$ with rate $R$ with our iterated interpolated protocol.
Before moving on, we note that we can, if we wish, instead say that, with probability $\Pr\left(Y_{\lceil N\left(R-\delta\right)\rceil}=1\right)$, we succeed in producing a state with fidelity at least $\Pr\left(Y_{\lceil N\left(R-\delta\right)\rceil}=1\right)$ to $W_{F'}^{\otimes\lceil N \left(R-\delta\right) \rceil}$, namely $\tilde{W}_{F'}^{\lceil N\left(R-\delta\right)\rceil}$, starting from $\rho^{\otimes N}$.

\subsubsection{Re-Parametrising the Interpolation} Now, for each $k \in \left\{1, 2, \dots, K\right\}$, let
\begin{equation}
q_{k} = p_{k}I_{k}
\end{equation}
so that the fidelity achieved is
\begin{equation}
\frac{q_{1}R_{1}F_{1}+q_{2}R_{2}F_{2}+\dots+q_{K}R_{K}F_{K}}{q_{1}R_{1}+q_{2}R_{2}+\dots+q_{K}R_{K}}
\end{equation}
and the rate is
\begin{equation}
\frac{q_{1}R_{1}+q_{2}R_{2}+\dots + q_{K}R_{K}}{q_{1}+q_{2}+\dots+q_{K}}\text{.}
\end{equation}
Note that the achievable fidelity and rate would not change if we multiplied $q_{1}, q_{2}, \dots, q_{K}$ by the same non-zero constant.
Therefore, in what follows, we will drop the normalisation condition that $p_{1}+p_{2}+\dots+p_{K} = 1$ and instead insist only that $q_{1}, q_{2}, \dots, q_{K}$ are non-negative and $q_{1}+q_{2}+\dots+q_{K} \neq 0$.

\subsubsection{Optimising the Interpolation} We now turn to the question of how to choose our $q_{1}, \dots, q_{K}$ so as to maximize the fidelity achieved at a particular rate or vice versa.
If one protocol achieved both at least as high a rate and at least as high a fidelity as another, then there would be no point in ever using the latter protocol.
Therefore, we assume $R_{1} > \dots > R_{K}$ and $F_{1} < \dots < F_{K}$.
Now, if at least $3$ of $q_{1}, q_{2}, \dots, q_{K}$ are positive, we can choose three distinct values, $i$, $j$ and $k$, from $\left\{1, 2, 3, \dots, K\right\}$ such that $q_{i}$, $q_{j}$ and $q_{k}$ are positive and assume without loss of generality that $i < j < k$.
Then, we can choose $r \in \left(0, 1\right)$ such that $rR_{i}+\left(1-r\right)R_{k} = R_{j}$.
\par
For every $l \in \left\{1, 2, \dots, K\right\} \setminus \left\{i, j, k\right\}$, let $\tilde{q}_{l} = q_{l}$.
If we have $rR_{i}F_{i}+\left(1-r\right)R_{k}F_{k}-R_{j}F_{j} \geq 0$, then we let $\tilde{q}_{i} = q_{i}+rq_{j}$, $\tilde{q}_{k} = q_{k} + \left(1-r\right)q_{j}$ and $\tilde{q}_{j} = 0\text{.}$
Otherwise, we let $\tilde{q}_{i} = q_{i} - rx$, $\tilde{q}_{k} = q_{k} - \left(1-r\right)x$ and $\tilde{q}_{j} = q_{j} + x$, where $x = \min{\left\{\frac{q_{i}}{r}, \frac{q_{k}}{1-r}\right\}}$.
In either case, we then see that $\tilde{q}_{1}, \tilde{q}_{2}, \dots, \tilde{q}_{K}$ are all non-negative, that the number of them that are non-zero is less than the number of $q_{1}, q_{2}, \dots, q_{K}$ that are non-zero, that
\begin{equation}\frac{\tilde{q}_{1}R_{1}+\dots + \tilde{q}_{K}R_{K}}{\tilde{q}_{1}+\dots+\tilde{q}_{K}} = \frac{q_{1}R_{1}+\dots + q_{K}R_{K}}{q_{1}+\dots+q_{K}}
\end{equation}
and that
\begin{equation}\frac{\sum_{k=1}^{K}{\tilde{q}_{k}R_{k}F_{k}}}{\sum_{k=1}^{K}{\tilde{q}_{k}R_{k}}} \geq \frac{\sum_{k=1}^{K}{q_{k}R_{k}F_{k}}}{\sum_{k=1}^{K}{q_{k}R_{k}}}\text{.}
\end{equation}
By repeating this argument, then, we see that there must be a way of interpolating between only two of Alice and Bob's protocols to achieve at least as high a Bell fidelity as did our original interpolated protocol at the same rate as our original interpolated protocol.
Therefore, when asking how we can optimize our choice of $q_{1}, q_{2}, \dots, q_{K}$ to maximize the Bell fidelity of the resulting states for some fixed rate or vice versa, we need only consider those choices of $q_{1}, q_{2}, \dots, q_{K}$ for which all but $2$ are non-zero.
This greatly simplifies the optimization problem.
\par
Suppose then that we decide to interpolate between two of Alice and Bob's protocols, say, the $i$th and $j$th protocols.
Suppose further that we are interested in producing approximate ebits at some rate, $R_{\text{target}}$ with $R_{i} \geq R_{\text{target}} \geq R_{j}$, with as high a Bell fidelity as possible.
We can then see that the fidelity we can achieve in this way is
\begin{equation}\frac{F_{i}R_{i}\left(R_{\text{target}} - R_{j}\right) + F_{j}R_{j}\left(R_{i} - R_{\text{target}}\right)}{R_{\text{target}}\left(R_{i}-R_{j}\right)}\text{.}
\end{equation}
Hence, provided $R_{K} \leq R_{\text{target}} \leq R_{1}$, the maximum fidelity we can achieve at our target rate by interpolating between the given entanglement distillation protocols is
\begin{equation}
\max_{\left(i, j\right) \in  A}{\frac{F_{i}R_{i}\left(R_{\text{target}} - R_{j}\right) + F_{j}R_{j}\left(R_{i} - R_{\text{target}}\right)}{R_{\text{target}}\left(R_{i}-R_{j}\right)}}\text{,}
\end{equation}
where $$A = \left\{\left(i, j\right) \in \left\{1, \dots, K\right\}^{2} \mid i < j, R_{\text{target}} \in \left[R_{j}, R_{i}\right]\right\}\text{.}$$
Likewise, the maximum rate at which we can produce approximate ebits with fidelity $F_{\text{target}} \in \left[F_{1}, F_{K}\right]$ by interpolating between the given entanglement distillation protocols is
\begin{equation}
\max_{\left(i, j\right) \in B}{\frac{\left(F_{j}-F_{i}\right)R_{i}R_{j}}{R_{i}\left(F_{\text{target}} - F_{i}\right)+R_{j}\left(F_{j} - F_{\text{target}}\right)}}\text{,}
\label{TargetFidelity}\end{equation}
where $$B = \left\{\left(i, j\right) \in \left\{1, \dots, K\right\}^{2} \mid i<j, F_{\text{target}} \in \left[F_{i}, F_{j}\right]\right\}\text{.}$$
\par
\begin{figure}
\vspace{-0.45cm}
\centering
\includegraphics[width=0.48\textwidth]{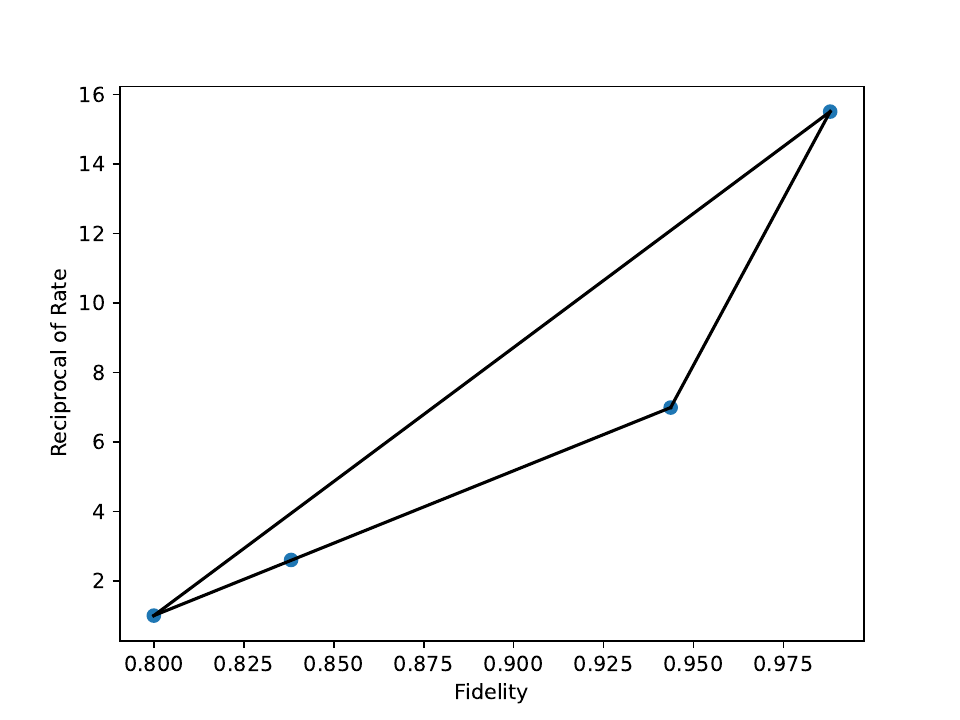}
\caption{The convex hull of four points, corresponding to the performance of $0$ iterations, $1$ iteration, $2$ iterations and $3$ iterations of the DEJMPS protocol applied to $W_{0.8}$ states. Note that, although it is unclear in the image, the point corresponding to $1$ iteration of the DEJMPS protocol lies strictly in the interior of the convex hull in this case.}
\label{fig:ConvexHull}
\end{figure}

Before moving on, we note that it is also possible to understand this optimization geometrically.
By Equations \ref{RateCalculation} and \ref{FidelityCalculation}, we can achieve some fidelity $F' \in \left(\frac{1}{2}, 1\right)$ at a rate $R \in \left(0, \infty\right)$ by interpolating between our protocols if and only if the point $\left(F, \frac{1}{R}\right)$ lies in the convex hull of the points $\left\{\left(F_{1}, I_{1}\right), \left(F_{2}, I_{2}\right), \dots, \left(F_{K}, I_{K}\right)\right\}$ plotted on the Cartesian plane.
Optimal interpolations correspond to points on the boundary of the convex hull.
An example is shown in Figure~\ref{fig:ConvexHull}.
From this geometrical perspective, we can see, just as we proved above, that we need not consider interpolations between more than $2$ protocols when searching for the optimal interpolation.
We can also see that, provided none of our protocols correspond to points in the strict interior of the convex hull, an optimal interpolation can always be found by interpolating between consecutive protocols in our list.

\subsubsection{Limiting the Number of Protocols} Suppose Alice and Bob have a sequence of protocols which achieve fidelities $F_{1}, F_{2}, \dots$ with respective rates $R_{1}, R_{2}, \dots$ with $F_{1} < F_{2} < \dots$, $R_{1} > R_{2} > \dots$, $\lim_{k \to \infty}{F_{k}} = 1$ and $\lim_{k \to \infty}{R_{k}} = 0$.
They want to find an interpolated protocol using the above ideas to, say, achieve a fidelity $F_{\text{target}}$ at as high a rate as possible.
However, without choosing a value of $K$, they cannot use Eq.~\eqref{TargetFidelity}.
How can they choose a value of $K$ and be sure that a larger value would not have allowed them to find an interpolated protocol that achieved the target fidelity at a higher rate?
\par
Naturally, we assume $F_{\text{target}} \in \left(F_{1}, 1\right)$.
Now, let $\tilde{K} = \max\left\{k \in \mathbb{N} \mid F_{k} < F_{\text{target}}\right\}$.
Suppose that Alice and Bob have already found some interpolated protocol which achieves the fidelity $F_{\text{target}}$ at a rate $R_{\text{achieved}} \in \left(0, \infty\right)$.
We will further assume that $R_{\tilde{K}+1} \leq R_{\text{achieved}} < R_{\tilde{K}}$ since we can clearly achieve fidelity $F_{\text{target}}$, which is less than or equal to $F_{\tilde{K}+1}$, with a rate at least as high as $R_{\tilde{K}+1}$ and if we could achieve it with a rate $R_{\tilde{K}}$ then the $\tilde{K}$th protocol would be strictly worse than some interpolation of two of our other protocols and could be omitted from our set of protocols without any loss in performance.
Now, if, for some $i \in \left\{1, 2, \dots, \tilde{K}\right\}$ and some $j \in \left\{\tilde{K}+1, \tilde{K}+2, \dots\right\}$, it is possible to achieve fidelity $F_{\text{target}}$ with rate greater than $R_{\text{achieved}}$ by interpolating between the $i$th and $j$th protocols, then, from Eq.~\eqref{TargetFidelity}, we see that
\begin{equation}
\frac{\left(F_{j}-F_{i}\right)R_{i}R_{j}}{R_{i}\left(F_{\text{target}}-F_{i}\right) + R_{j}\left(F_{j}-F_{\text{target}}\right)} > R_{\text{achieved}},
\end{equation}
and so we have
\begin{align}
    R_{j} &> \frac{\left(F_{\text{target}}-F_{i}\right)R_{\text{achieved}}R_{i}}{\left(F_{j}-F_{i}\right)R_{i} - \left(F_{j}-F_{\text{target}}\right)R_{\text{achieved}}} \nonumber \\
    &> \frac{\left(F_{\text{target}}-F_{i}\right)R_{\text{achieved}}R_{i}}{\left(1-F_{i}\right)R_{i} - \left(1-F_{\text{target}}\right)R_{\text{achieved}}} 
    \nonumber
    \\
    &\geq \Omega,
\end{align}
where $\Omega$ denotes
\begin{equation}\label{MaximumSet}
\min_{i' \in \left\{1, \dots, \tilde{K}\right\}}{\frac{\left(F_{\text{target}}-F_{i'}\right)R_{\text{achieved}}R_{i'}}{\left(1-F_{i'}\right)R_{i'} - \left(1-F_{\text{target}}\right)R_{\text{achieved}}}}\text{.}
\end{equation}
Note that the expression in Eq.~\ref{MaximumSet} is positive and independent of $i$ and $j$. Therefore, Alice and Bob can set $K$ to  $\max{\left\{k \in \mathbb{N} \mid R_{k} > \Omega\right\}}$, thereby restricting their attention to a finite number of protocols to interpolate between while knowing that they could not have achieved a better rate by setting $K$ to a larger value.
We can use an almost identical formula to limit $K$ if we want to maximize the fidelity we achieve at a particular fixed rate.

\subsection{The DEJMPS Case}

\subsubsection{DEJMPS Interpolation}\label{DEJMPSInterpolation} We will now apply these concepts to a specific example.
Imagine Alice and Bob want to produce approximate ebits with Bell fidelity $F_{\text{target}} \in \left(\frac{1}{2}, 1\right)$ at the highest rate they can by interpolating between DEJMPS protocols with different numbers of iterations.
For a description of the DEJMPS protocol, see Appendix~\ref{DEJMPSDescription}.
\par
Let $s_{0} = 1$ and, for $i \in \mathbb{N}$, let $s_{i} = t_{i}s_{i-1} = \prod_{j=1}^{i}t_{j}$, where $t_{i}$ is the probability of the DEJMPS protocol succeeding when applied to two independent pairs, each in the output state of applying the DEJMPS protocol $i-1$ times.
If we were applying successive iterations of the DEJMPS protocol to a large pool of states, then, for every $i \in \mathbb{N}$, the fraction of the qubit pairs we would expect to survive the $i$th iteration of the DEJMPS protocol would be $\frac{t_{i}}{2}$.
Then we can see that, for $i \in \left\{0, 1, 2, \dots\right\}$, the rate of the DEJMPS protocol performed $i$ times is $R_{i} = \frac{s_{i}}{2^{i}}$.
For an alternative derivation of this result, see Appendix~\ref{DEJMPSAsymptoticApproximation}.
Figure~\ref{fig:DEJMPSOnly} shows the rates at which the Werner states produced by applying a depolarizing channel to a true ebit can be converted into states with fidelity $0.9$ to a true ebit via interpolated protocols of this kind.

\begin{figure}
\vspace{-0.4cm}
\centering
\includegraphics[width=0.48\textwidth]{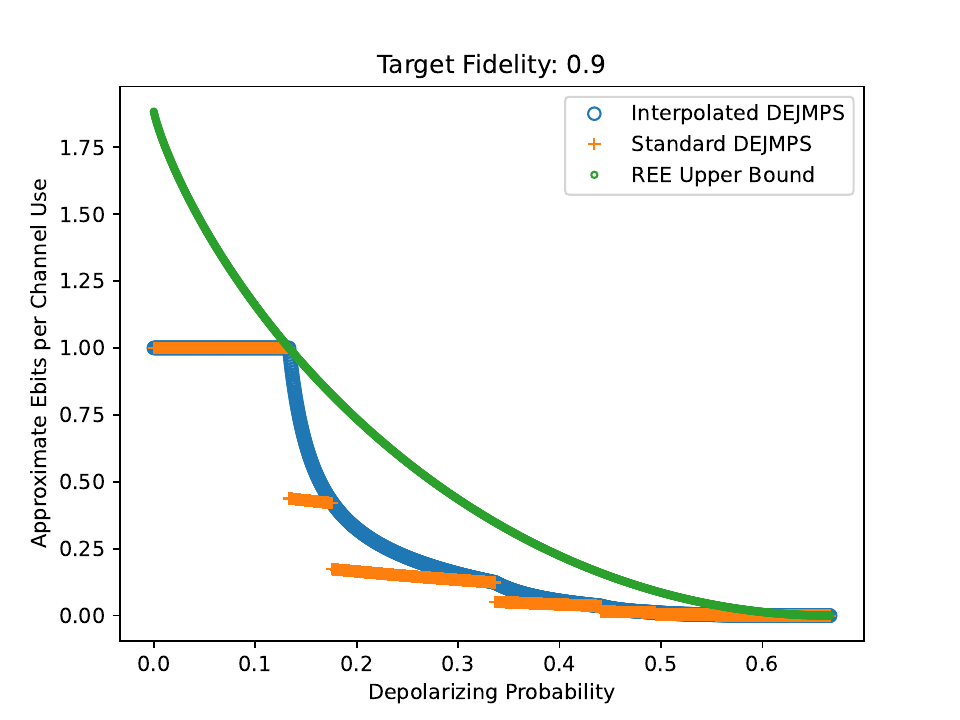}
\caption{The rates at which we can convert the output states of the depolarizing channel acting on true ebits into approximate ebits with fidelity $0.9$ to a true ebit using the DEJMPS and interpolated DEJMPS protocols. The rate achievable by interpolating between different numbers of DEJMPS iterations is shown in blue. The rate achievable by simply applying the DEJMPS protocol iteratively without interpolation until we reach the target fidelity is shown in orange. The upper bound on the achievable rate shown in green is based on the relative entropy of entanglement (REE).}
\label{fig:DEJMPSOnly}
\end{figure}

\subsubsection{An Upper Bound on the Achievable Rate} For reference, it may also be useful to compare the rate at which the standard, iterated DEJMPS and interpolated DEJMPS protocols achieve our target fidelity to some upper bound on the achievable rate.
Suppose a fidelity $F_{\text{target}}$ is achievable with rate $R$ from $\rho$.
Now, let $E$ be an asymptotically continuous entanglement measure.
We can see then that, for all $\delta \in \left(0, R\right)$, we must have 
\begin{equation}
    \lim_{N \to \infty}\frac{E\left(W_{F_{\text{target}}}^{\otimes\lceil N\left(R-\delta\right) \rceil}\right)}{N} \leq \lim_{N \to \infty}\frac{E\left(\rho^{\otimes N}\right)}{N}\text{.}
\end{equation}
For example, suppose $E$ is the relative entropy of entanglement (REE) and $\rho$ is some Werner state, $W_{F_{\text{initial}}}$.
Although it is not additive in general, it is additive with respect to identical Werner states and so we find~\cite{RelativeEntropyOfEntanglement, AsymptoticRelativeEntropyOfEntanglement} 
\begin{equation}
    R \leq \frac{1-h(F_{\text{initial}})}{1-h(F_{\text{target}})}\text{,}
\end{equation}
where $h$ is the binary entropy function. See again Fig.~\ref{fig:DEJMPSOnly}, where we compare this upper bound with standard and interpolated DEJMPS protocols. 

\section{The Finite-Size Scenario}\label{Finite-Size}

\subsection{Finite $N$ Performance}
In practice, the pool of states on which we perform entanglement purification will be finite.
Therefore, it would be useful to be able to quantify how close the protocols described in Section~\ref{AsymptoticInterpolation} come to achieving their asymptotic fidelity and rate in the case where we are limited to a pool of $N$ initial states.
As before, we let $F'$ denote the fidelity our protocol achieves in the limit as $N \to \infty$.
If we aim to produce $M'$ states and the joint output state for our protocol acting on $\rho^{\otimes N}$ is $\sigma^{M'}$, then we can define the global infidelity as
\begin{equation}
    \epsilon_{\text{global}, M'} = 1 - F\left(\sigma^{M'}, W_{F'}^{\otimes M'}\right)
\end{equation}
and the (average) per-pair infidelity as
\begin{equation}
    \epsilon_{\text{pair}, M'} = 1 - \left(1 - \epsilon_{\text{global}, M'}\right)^{\frac{1}{M'}}\text{.}
\end{equation}
Here, we will be interested in the maximum number of states, $M$, that Alice and Bob can aim to produce while keeping either the global infidelity or the per-pair infidelity below some chosen threshold, $\tilde{\epsilon}$, in $\left(0, 1\right)$.
That is, we either set
\begin{equation}\label{GlobalDefinition}
    M = \max{\left\{M' \in \left\{0, \dots, N\right\} \mid \epsilon_{\text{global}, M'} \leq \tilde{\epsilon}\right\}}
\end{equation}
or
\begin{equation}\label{PairDefinition}
    M = \max{\left\{M' \in \left\{0, \dots, N\right\} \mid \epsilon_{\text{pair}, M'} \leq \tilde{\epsilon}\right\}}\text{,}
\end{equation}
depending on whether we are interested in the global or the per-pair infidelity.
We will then say that, from $\rho^{\otimes N}$, we have achieved a fidelity $F'$ and rate $\frac{M}{N}$ with a global or per-pair infidelity at most $\tilde{\epsilon}$.
\par
We reiterate at this point that, although, for convenience, we will speak of a pool of $N$ initial states, these states need not all exist simultaneously.
For instance, suppose two nodes in a quantum network generate initial entangled states, $\rho$, apply their protocol as much as they can to what has been generated so far and immediately use their purified pairs as they produce them.
If, for example, they are using an interpolated DEJMPS protocol, as described in Section~\ref{DEJMPSInterpolation}, in which the largest number of iterations of the DEJMPS protocol they apply is $j$, they need only store up to $j + 1$ qubit pairs at a time. 
Then, if we know that, say, in some time period, the two nodes will generate $N$ copies of $\rho$ and we want to purify them into as many copies of $W_{F'}$ as possible, up to some specified infidelity, then we can achieve this with some number of purified states, $M$, even though the copies of $\rho$ need not all exist simultaneously.

\begin{figure*}[ht]
\centering
\includegraphics[width=\textwidth]{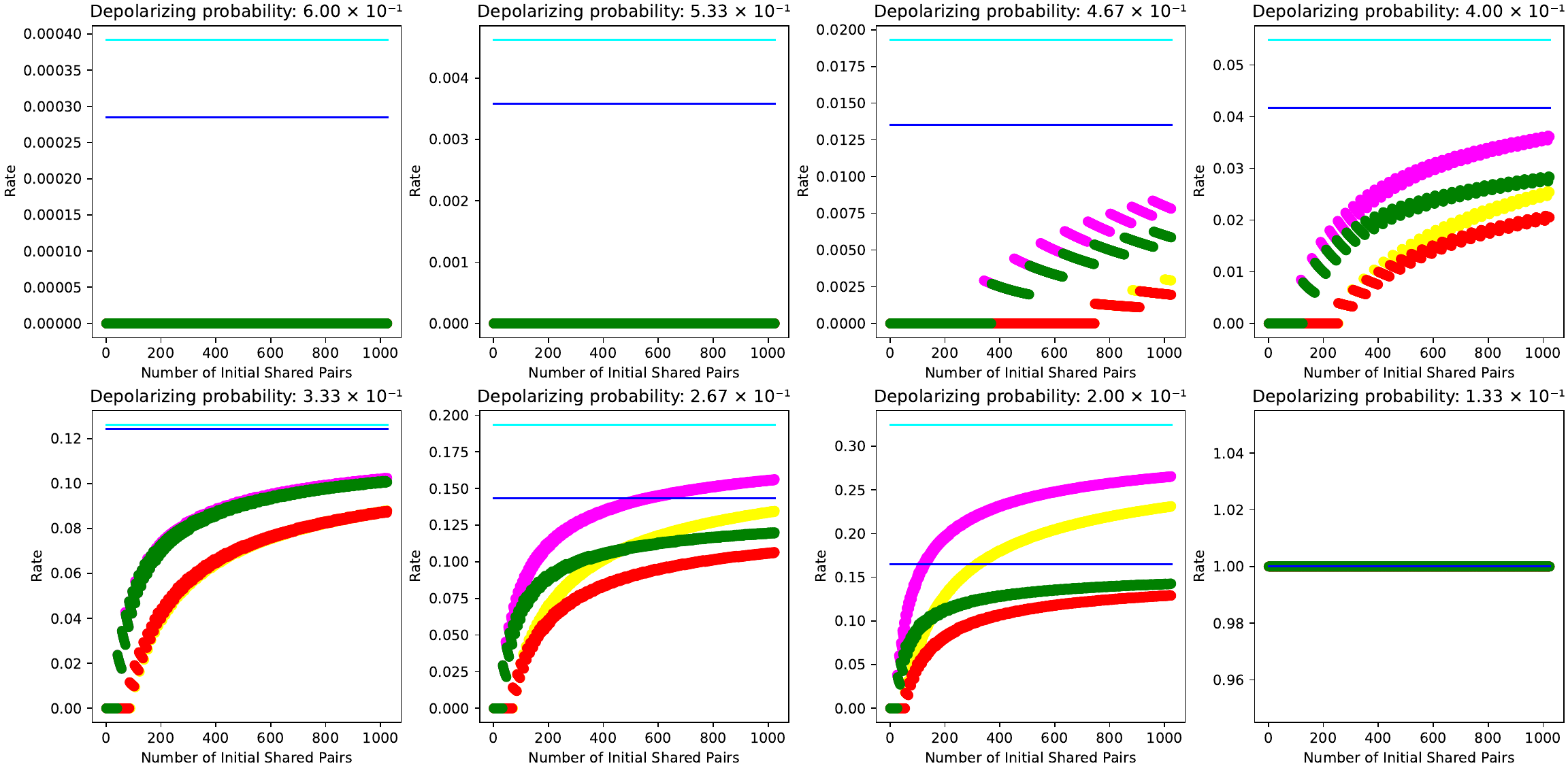}
\caption{Lower and upper bounds on the maximum number of output pairs per input pair at which the DEJMPS and interpolated DEJMPS protocols achieve the target fidelity of $0.9$ within a global infidelity of $10^{-7}$ against the number of initial pairs, generated by applying a depolarizing channel to one qubit of our chosen Bell state. The Bell fidelities of the initial states are linearly spaced, ranging from $0.55$ in the top-left figure to $0.9$ in the bottom-right. At the top of each panel we show the depolarizing probabilities corresponding to those fidelities to three significant figures. If the depolarizing probability is $p$, then the Bell fidelity of the initial states is $\frac{4 - 3p}{4}$. The lower and upper bounds for the DEJMPS protocol are in red and green, respectively, and for the interpolated DEJMPS protocol in yellow and magenta, respectively. The asymptotic rates of the DEJMPS and interpolated DEJMPS protocols are shown in blue and cyan, respectively. Note that the lower bounds for the interpolated protocols are based on Inequality~\ref{InterpolatedLowerBound}, the lower bounds for the uninterpolated DEJMPS protocols are based on Inequality~\ref{StandardLowerBound} and the upper bounds are based on Inequality~\ref{UpperBound}.}
\label{fig:FiniteSizeDepolarising}
\end{figure*}

\begin{figure*}[ht]
\centering
\includegraphics[width=\textwidth]{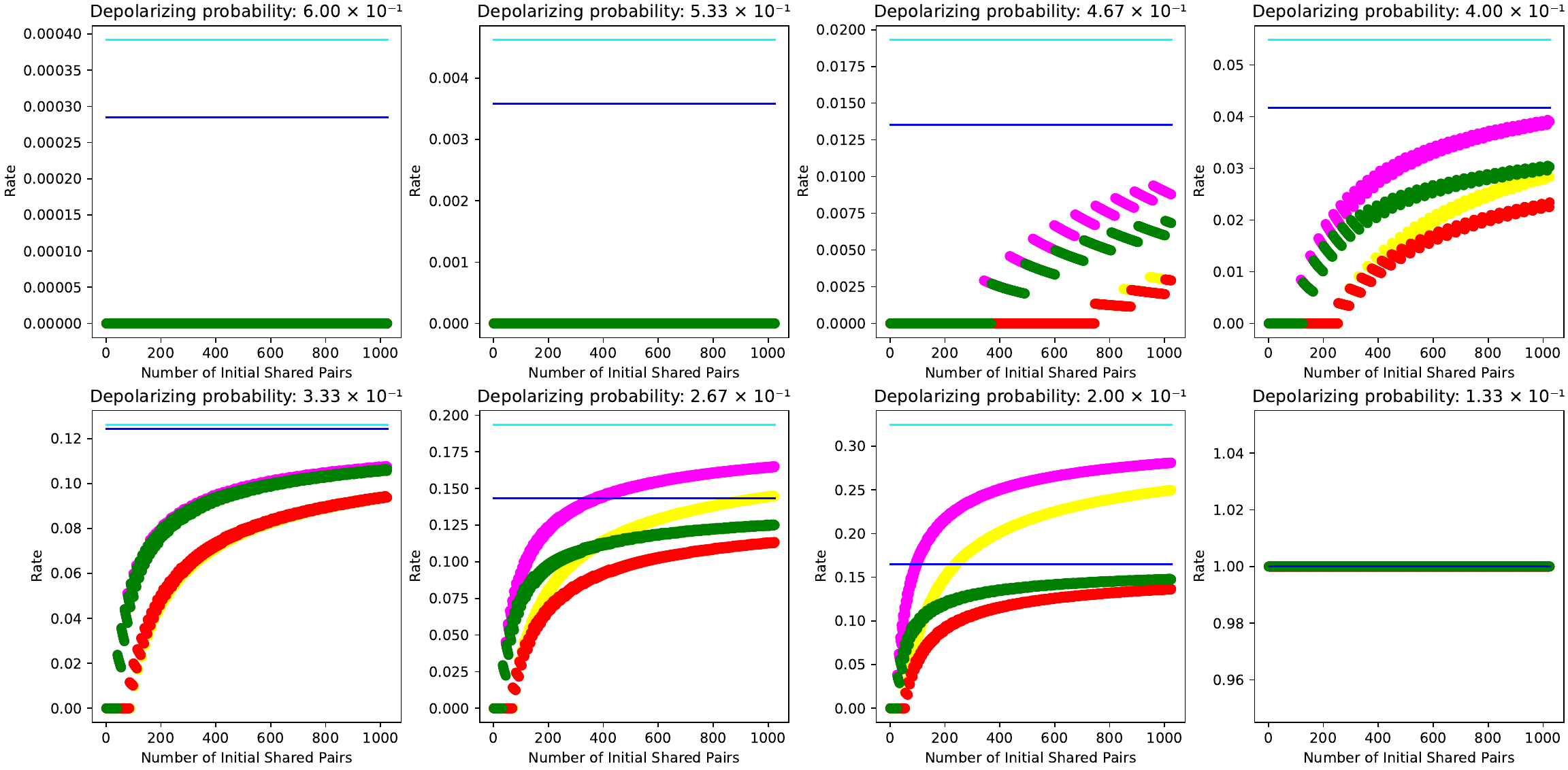}
\caption{The same as Figure~\ref{fig:FiniteSizeDepolarising}, except the maximum allowed infidelity is a per-pair infidelity of $10^{-7}$.}
\label{fig:FiniteSizeDepolarisingPerPair}
\end{figure*}

\subsection{Bounding $M$}
In practice, it may be computationally expensive to find $M$.
Therefore, we will find upper and lower bounds on it instead by placing upper and lower bounds on $F\left(\sigma^{M'}, W_{F'}^{\otimes M}\right)$ for different values of $M'$.
Inequality~\eqref{FidelityLowerBound} tells us that we will have $$F\left(\sigma^{M'}, W_{F'}^{\otimes M'}\right) \geq \Pr\left(Y_{M'}=1\right)^{2}\text{.}$$
If we do not use an interpolated protocol but rather one which always produces independent $W_{F'}$ states, then we can use the independence of $Z^{M'}$ and $Y_{M'}$ to modify the argument used to derive Inequality~\eqref{FidelityLowerBound} and obtain a tighter lower bound:
\begin{align}
    &F\left(\sigma^{M'}, W_{F'}^{\otimes M'}\right) \nonumber \\
    \geq &\left(\sum_{m\in\left\{1, 2, 3, 4\right\}^{M'}}{\sqrt{\Pr\left(Z^{M'} = z\right)^{2} \Pr\left(Y_{M'} = 1\right)}}\right)^{2} \nonumber \\
    = &\Pr\left(Y_{M'}=1\right)\text{.} \label{UninterpolatedLowerBound}
\end{align}
For instance, Inequality~\eqref{UninterpolatedLowerBound} would apply if our protocol consisted in applying enough iterations of the DEJMPS protocol for the Bell fidelity of the resulting output states to exceed $F'$ and then applying a twirling and a depolarizing operation to convert our output states into $W_{F'}$ states.
We can obtain an upper bound on $F\left(\sigma^{M'}, W_{F'}^{\otimes M'}\right)$ by using the fact that the fidelity between two states increases under the action of the partial trace on any of the subsystems.
Let $F_{M'}''$ equal
\begin{equation}
\frac{\Pr\left(Y_{M'}=0\right)}{2} + \sum_{k = 1}^{K}{\Pr\left(X_{M'}=k, Y_{M'}=1\right)}F_{k}\text{.}
\end{equation}
Then, taking the partial trace over all but the last qubit pair, we find that
\begin{align}
&F\left(\sigma^{M'}, W_{F'}^{\otimes M'}\right) \nonumber \\
\leq &F\left(W_{F''}, W_{F'}\right) \nonumber \\
= &\left(\sqrt{F'F_{M'}''}+\sqrt{\left(1-F'\right)\left(1-F_{M'}''\right)}\right)^{2}\text{.}\label{TraceUpperBound}
\end{align}
\par
We will use these upper and lower bounds to put upper and lower bounds on $M$, allowing us to compare the performance of our new, interpolated protocols to our original protocols.
For concreteness, we will assume that we want to limit the global infidelity of our protocol's output to be less than $\tilde{\epsilon}$, that is we are defining $M$ using Eq.~\eqref{GlobalDefinition}.
If we instead wish to use Eq.~\eqref{PairDefinition}, what follows can be adapted straightforwardly to that case.
\par
We have
\begin{equation}
    M \geq \max{O_{1}}\text{,}\label{InterpolatedLowerBound}
\end{equation}
where $O_{1}$ denotes
\begin{equation}
\left\{M' \in \left\{0, \dots, N\right\} \mid \Pr\left(Y_{M'}=1\right) \geq \sqrt{1 - \tilde{\epsilon}}\right\}\text{.}
\end{equation}
If our protocol only produces independent $W_{F'}$ states upon success, we can improve this using Inequality~\ref{UninterpolatedLowerBound} and write
\begin{equation}
    M \geq \max{O_{2}}\text{,}\label{StandardLowerBound}
\end{equation}
where
\begin{equation}
       O_{2} = \left\{M' \in \left\{0, \dots, N\right\} \mid \Pr\left(Y_{M'}=1\right) \geq 1 - \tilde{\epsilon}\right\}\text{.}
\end{equation}
Finally, based on Inequality~\ref{TraceUpperBound}, we have
\begin{equation}
    M \leq \max{O_{3}}\text{,}\label{UpperBound}
\end{equation}
where $O_{3}$ denotes
\begin{equation}
\left\{M' \in \left\{0, \dots, N\right\} \mid F\left(W_{F'}, W_{F_{M'}''}\right) \geq 1 - \tilde{\epsilon}\right\}\text{.}
\end{equation}
Hence, if we have a method for calculating $\Pr\left(X_{M'} = k, Y_{M'}=1\right)$ for different values of $M'$ and $k$, then we can bound $M$. In Appendix~\ref{ExactCalculation}, we describe two such methods that we have applied to our interpolated DEJMPS protocols to produce Figures \ref{fig:FiniteSizeDepolarising} and \ref{fig:FiniteSizeDepolarisingPerPair}.
These figures provide an example of the improved performance that can be obtained in certain regimes using the interpolation techniques described in this paper. In particular, we see how the interpolated DEJMPS protocol is able to clearly outperform the rate of the standard one in various intermediate regimes. 
Appendix~\ref{ExtraPlots} provides further such figures.

\section{Conclusion}

Interpolating between different protocols often allows two parties to purify weakly entangled pairs into pairs with a given target Bell fidelity at a higher rate or to purify at a particular rate input pairs into output pairs with a higher Bell fidelity than could have been done by any of their ``uninterpolated'' protocols.
This is particularly true when, as will typically be the case when choosing between a discrete set of protocols, none of our protocols achieve the target fidelity or target rate without excess.
Therefore, interpolating between entanglement purification protocols is likely to be a useful technique for any application for which we require pairs to be purified at a particular rate or to a particular quality.
\par
For simplicity, we have only considered protocols that purify just one pair at a time, so called $n$-\textgreater$1$ protocols, here but our results could be extended to protocols that produce multiple purified pairs together, so called $n$-\textgreater$m$ protocols.
When we use $n$-\textgreater$m$ protocols, the situation is complicated by the fact that the purified pairs are no longer independent but are correlated with pairs that were produced with them.
For instance, if all our protocols produce two Werner states of a particular Bell fidelity, rather than one, then if you know what the state of the first purified pair is, you will also know the state of the second.
For some applications, these correlations between states may not matter but for others they might.
To produce a global state which is approximately equal to a tensor product of Werner states with the target fidelity, as measured by the fidelity of our average global output state to that global target state, we could purify many pairs and then perform a random permutation of the purified pairs.
One disadvantage of this would be that we would not be able to use the purified pairs as soon as they were purified.
Rather, we would have to wait until we had purified many pairs so we could perform our permutation.
\par
One question not considered here is how to interpolate optimally in the finite-size scenario.
The interpolation that is optimal in the asymptotic scenario is not necessarily optimal when we have a finite $N$, as can be seen, for example, in Figure~\ref{fig:FiniteSizeDepolarising}, where the standard DEJMPS protocol performs better than our interpolated protocol for certain channel parameters and values of $N$.
This could be an interesting avenue for future research.
\par
A natural next step would be to apply the results developed here in a realistic setting.
For instance, it would be of interest to model how effectively they can be applied as part of a practical entanglement distribution protocol in a realistic quantum network~\cite{Pirandola2019,Shchukin2019,Lopiparo2020,Harney2022,Harney2022b,Lopiparo2025}.
Note that, while we have chosen to use the DEJMPS protocol to illustrate the interpolation idea, when designing an entanglement distribution protocol, we may be able to achieve better performance by considering other practical entanglement purification protocols.

\section*{Acknowledgments}

\noindent We acknowledge support from UKRI via the ``Integrated Quantum Networks Research Hub'' (IQN, EP/Z533208/1).
We would like to thank Cillian Harney for insightful discussions and the anonymous referees for their helpful comments.

\appendix

\section{The DEJMPS Protocol}\label{DEJMPSDescription}

\subsection{The Basic Description}

The DEJMPS protocol \cite{DEJMPS} is a well-known entanglement purification protocol.
It is very similar to another well-known purification protocol, often referred to as the ``BBPSSW" protocol \cite{BBPSSW}, although the DEJMPS protocol is regarded as more efficient \cite{PurificationReview}.
As we have used the DEJMPS protocol to illustrate the interpolation techniques developed in this paper, we will briefly describe it here.
\par
Let $\Ket{0_{A}}$ and $\Ket{1_{A}}$ denote the two computational basis states for Alice's qubit and $\Ket{0_{B}}$ and $\Ket{1_{B}}$ denote the same for Bob's qubit.
Now, let
\begin{equation}
    \Ket{\Phi^{+}} = \frac{\Ket{0_{A}}\Ket{0_{B}}+\Ket{1_{A}}\Ket{1_{B}}}{\sqrt{2}}\text{,}
\end{equation}
\begin{equation}
    \Ket{\Phi^{-}} = \frac{\Ket{0_{A}}\Ket{0_{B}}-\Ket{1_{A}}\Ket{1_{B}}}{\sqrt{2}}\text{,}
\end{equation}
\begin{equation}
    \Ket{\Psi^{+}} = \frac{\Ket{0_{A}}\Ket{1_{B}}+\Ket{0_{A}}\Ket{1_{B}}}{\sqrt{2}}
\end{equation}
and
\begin{equation}
    \Ket{\Psi^{-}} = \frac{\Ket{0_{A}}\Ket{1_{B}}-\Ket{0_{A}}\Ket{1_{B}}}{\sqrt{2}}\text{.}
\end{equation}
We will let $\Ket{\Phi^{+}}$ be our target Bell state, which we referred to simply as $\Ket{\Phi}$ in the main body of the paper. Note that we may repeat the argument for any other target Bell state.
\par
Suppose that Alice and Bob share two qubit pairs in the joint state $\rho \otimes \rho'$, where $\rho$ is the state of the first pair and $\rho'$ that of the second pair.
If $C$ denotes either $A$ or $B$, let $U_{C}$ be the unitary operation that acts on one of Alice or Bob's qubits accordingly and satisfies
\begin{equation}
    U_{C}\Ket{0_{C}} = \frac{\Ket{0_{C}}-i\Ket{1_{C}}}{\sqrt{2}}
\end{equation}
and
\begin{equation}
    U_{C}\Ket{1_{C}} = \frac{\Ket{1_{C}}-i\Ket{0_{C}}}{\sqrt{2}}\text{.}
\end{equation}
Now, Alice applies $U_{A}$ to both of her qubits and Bob applies $U_{B}^{\dagger}$ to both of his.
Alice and Bob then each perform a Controlled-Not operation on their respective qubits, with the qubits from the first pair being the source qubits and the qubits from the second pair being the target qubits.
Finally, Alice and Bob measure their target qubits in the computational basis.
If their measurement results are the same, they keep the first qubit pair and otherwise they discard them.
\par
Let $A$, $B$, $C$ and $D$ denote the diagonal elements of $\rho$ in the Bell basis in the order $\Ket{\Phi^{+}}$, $\Ket{\Psi^{-}}$, $\Ket{\Psi^{+}}$, $\Ket{\Phi^{-}}$ and let $A'$, $B'$, $C'$ and $D'$ denote the corresponding terms for the state $\rho'$.
In that case, the probability that the protocol succeeds and we keep the source pair is
\begin{equation}
    t = \left(A+B\right)\left(A'+B'\right) + \left(C+D\right)\left(C'+D'\right)\text{.}
\end{equation}
If the protocol succeeds, the diagonal elements of the resulting state in the Bell basis with the same ordering as before are given by
\begin{equation}
    A'' = \frac{AA'+BB'}{t}\text{,}
\end{equation}
\begin{equation}
    B'' = \frac{CD'+C'D}{t}\text{,}
\end{equation}
\begin{equation}
    C'' = \frac{CC'+DD'}{t}
\end{equation}
and
\begin{equation}
    D'' = \frac{AB'+A'B}{t}\text{.}
\end{equation}

\subsection{Iterations of the DEJMPS Protocol}

In the main body of the paper, we speak of performing multiple iterations of the DEJMPS protocol.
Here, we clarify what we mean by this.
Suppose we start with a large pool of independent qubit pairs in some weakly entangled state, $\rho$.
\par
\begin{figure}
\centering
\includegraphics[width=0.5\textwidth]{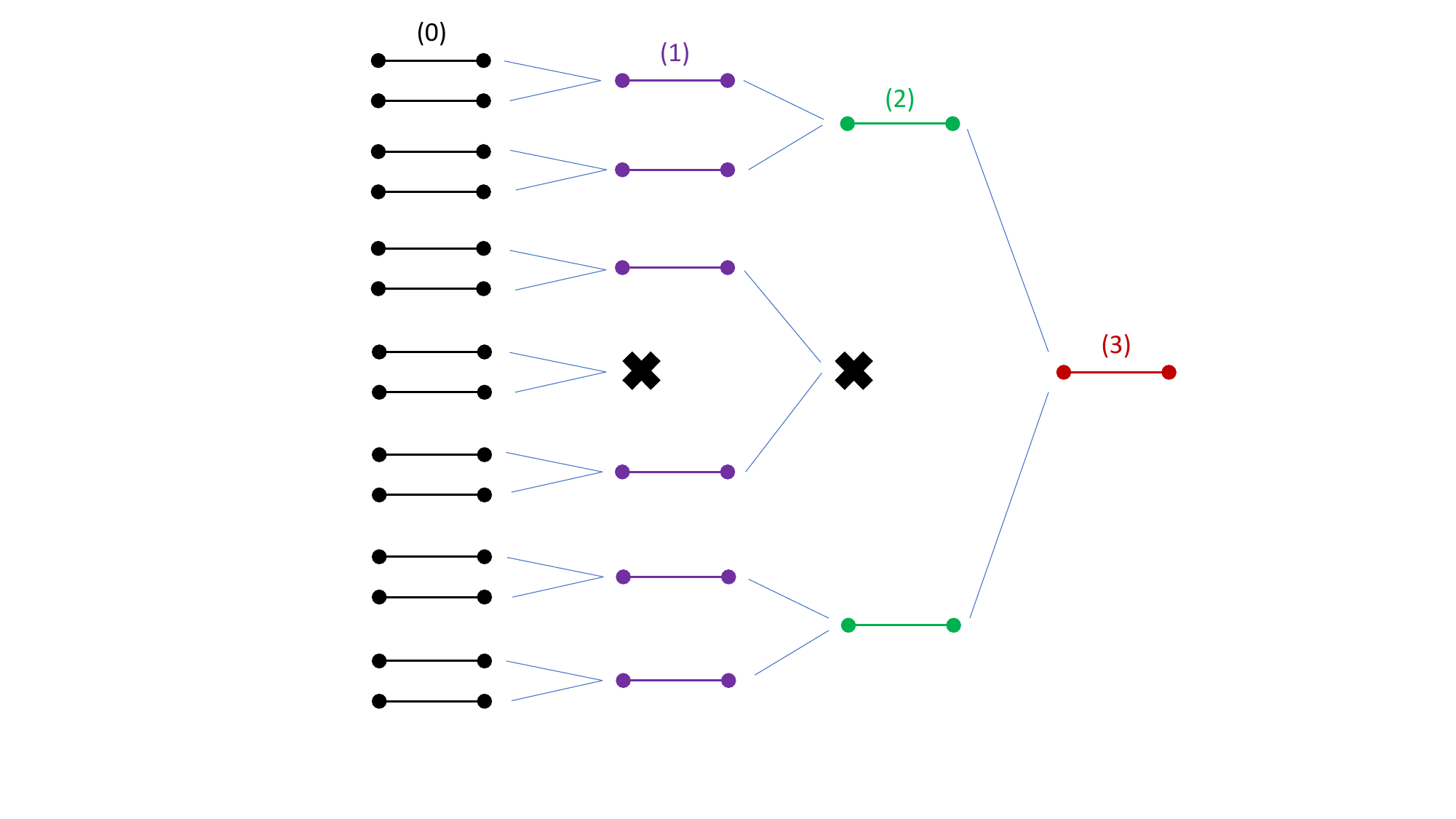}
\vspace{-0.6cm}
\caption{An example of what performing $3$ iterations of the DEJMPS protocol might look like. The black lines represent the initial pairs. The purple lines represent pairs produced via $1$ iteration of the DEJMPS protocol. The green lines represent pairs produced via $2$ iterations of the DEJMPS protocol. The red line represents a pair produced via $3$ iterations of the DEJMPS protocol. The black crosses denote failed applications of the protocol.}
\label{fig:DEJMPSIterations}
\end{figure}

Performing $0$ iterations of the DEJMPS protocol means leaving the initial state alone.
Performing $1$ iteration of the DEJMPS protocol means repeatedly applying the standard DEJMPS protocol to pairs of qubits until we succeed in producing a more entangled pair.
Performing $2$ iterations of the DEJMPS protocol means applying $1$ iteration of the DEJMPS protocol two times to produce two pairs more entangled than the initial pairs and then applying the DEJMPS protocol to the resulting two pairs to produce an even more entangled pair or to fail, in which case we repeat the whole procedure until we succeed in producing our even more entangled pair.
This pattern then continues.
An example of what applying $3$ iterations of the DEJMPS protocol might look like is shown in Figure~\ref{fig:DEJMPSIterations}.
Whenever we speak of applying the DEJMPS protocol or some other purification protocol until our states reach the target fidelity, we are talking about applying Algorithm~\ref{alg:TargetFidelity}.
Algorithm~\ref{alg:DEJMPSTargetFidelity} is equivalent to Algorithm~\ref{alg:TargetFidelity} when we are applying the DEJMPS protocol.

\begin{algorithm}
\caption{Applying an entanglement purification protocol until we reach a target Bell fidelity, $F'$. $P\left(\rho, N\right)$ denotes the probability distribution for the number of pairs we will purify if we apply the protocol once, $Q\left(\rho\right)$ denotes the states of the purified pairs after one successful application of the protocol and $\sim$ denotes sampling from a probability distribution.}\label{alg:TargetFidelity}
\begin{algorithmic}
\While{$\Bra{\Phi}\rho \Ket{\Phi} < F'$}
    \State $N \sim P\left(\rho, N\right)$
    \State $\rho \gets Q\left(\rho\right)$
\EndWhile
\State $\rho \gets W_{F'}$
\end{algorithmic}
\end{algorithm}

\begin{algorithm}
\caption{An equivalent alternative to Algorithm~\ref{alg:TargetFidelity} in the case where we are applying the DEJMPS protocol. We adopt the notation of Algorithm~\ref{alg:TargetFidelity} and use $\mathrm{B}\left(\lfloor \frac{N}{2}\rfloor, t\right)$ to denote a binomial random variable with $\lfloor \frac{N}{2}\rfloor$ trials and a trial success probability of $t$.}\label{alg:DEJMPSTargetFidelity}
\begin{algorithmic}
\State $A \gets \Bra{\Phi^{+}}\rho \Ket{\Phi^{+}}$
\State $B \gets \Bra{\Psi^{-}}\rho \Ket{\Psi^{-}}$
\State $C \gets \Bra{\Psi^{+}}\rho \Ket{\Psi^{+}}$
\State $D \gets \Bra{\Phi^{-}}\rho \Ket{\Phi^{-}}$
\While{$A < F'$}
    \State $t \gets \left(A+B\right)^{2} + \left(C+D\right)^{2}$
    \State $N \sim \mathrm{B}\left(\lfloor \frac{N}{2} \rfloor, t\right)$
    \State $A' \gets \frac{A^{2} + B^{2}}{t}$
    \State $B' \gets \frac{2CD}{t}$
    \State $C' \gets \frac{C^{2} + D^{2}}{t}$
    \State $D' \gets \frac{2AB}{t}$
    \State $A \gets A'$
    \State $B \gets B'$
    \State $C \gets C'$
    \State $D \gets D'$
\EndWhile
\State $\rho \gets W_{F'}$
\end{algorithmic}
\end{algorithm}

\subsection{Pairs Consumed and Pairs Produced}

\subsubsection{Pairs Produced from Fixed Pool}

In the main body of the paper, we measure the error of our purification protocols by the infidelity between the average output state of the protocol and the desired number of copies of the target Werner state.
We do this because it allows us to compare our interpolated protocols to our uninterpolated protocols.
However, when simply performing a fixed number of iterations of the DEJMPS protocol to a pool of independent and identical initial states, the only variable is the number of output pairs we successfully purify.
For completeness, we show one way in which the distribution of this random variable may be calculated.
\par
For $\left(k, n\right) \in \left\{0, 1, 2, \dots\right\}^{2}$, let $M_{k}^{n}$ denote the random variable representing the number of pairs we can purify from $k$ iterations of the DEJMPS protocol from $n$ initial pairs.
Likewise, for all $k \in \mathbb{N}$, let $t_{k}$ denote the probability that the DEJMPS protocol will succeed when applied to two pairs produced via $k-1$ iterations of the protocol.
Then, for all $\left(m, n\right) \in \left\{0, 1, 2, \dots\right\}^{2}$, we have
\begin{equation}
\Pr\left(M_{0}^{n} = m\right) = \delta_{m, n}\text{,}
\end{equation}
where $\delta$ is the Kronecker delta, and
\begin{equation}
\mathbb{E}\left(M_{0}^{n}\right) = n\text{.}
\end{equation}
\par
Similarly, we can see that, for all $n \in \left\{0, 1, 2, \dots\right\}$, $M_{1}^{n}$ follows a binomial distribution.
For every $t \in \left[0, 1\right]$ and $n \in \left\{0, 1, 2, \dots\right\}$, let $B_{t}^{n}$ be a $\mathrm{B}\left(n, t\right)$ random variable.
Then, for all $m \in \left\{0, 1, 2, \dots\right\}$,
\begin{equation}
\Pr\left(M_{1}^{n} = m\right) = \Pr\left(B_{t_{1}}^{\lfloor \frac{n}{2} \rfloor} = m\right)
\end{equation}
and
\begin{equation}
\mathbb{E}\left(M_{1}^{n}\right) = \lfloor \frac{n}{2} \rfloor t_{1}\text{.}
\end{equation}
\par
Now, we can derive a recurrence relation relating the distribution of $M_{k}^{n}$ to the distribution of $M_{k-1}^{n}$ by conditioning on how many pairs we will be able to purify after $k-1$ iterations of the protocol.
For all $k \in \mathbb{N}$ and $\left(m, n\right) \in \left\{0, 1, 2\dots \right\}^{2}$, by conditioning on the number of pairs distilled after $k-1$ iterations, we find that $\Pr\left(M_{k}^{n} = m\right)$ is given by
\begin{equation}
\sum_{m_{k-1} = 0}^{\lfloor \frac{n}{2^{k-1}} \rfloor}{\Pr\left(M_{k-1}^{n}=m_{k-1}\right)\Pr\left(B_{t_{k}}^{\lfloor \frac{m_{k-1}}{2} \rfloor} = m\right)}
\end{equation}
and
\begin{align}\label{ExpectationRecurrence}
    \mathbb{E}\left(M_{k}^{n}\right) &= \sum_{m_{k-1} = 0}^{\lfloor \frac{n}{2^{k-1}} \rfloor}\Pr\left(M_{k-1}^{n}=m_{k-1}\right)\lfloor \frac{m_{k-1}}{2} \rfloor t_{k} \nonumber \\
    &= \mathbb{E}\left(\lfloor \frac{M_{k-1}^{n}}{2} \rfloor\right)t_{k} \nonumber \\
    &= \frac{t_{k}\left(\mathbb{E}\left(M_{k-1}^{n}\right) - \Pr\left(M_{k-1}^{n} \in O\right)\right)}{2}\text{,}
\end{align}
where $O$ is the set of positive odd integers.
\par
As an example, let us use Eq.~\eqref{ExpectationRecurrence} to find $\mathbb{E}\left(M_{2}^{n}\right)$ for all $n \in \left\{0, 1, 2\dots\right\}$.
We see that, for all $n \in \left\{0, 1, 2, \dots \right\}$,
\begin{align}
    \Pr\left(M_{1}^{n} \in O\right) &= \sum_{m_{1} \in \left\{1, 3, 5, \dots \right\}}{\binom{\lfloor \frac{n}{2} \rfloor}{m_{1}}t_{1}^{m_{1}}\left(1-t_{1}\right)^{\lfloor \frac{n}{2} \rfloor - m_{1}}} \nonumber \\
    &= \frac{1 - \left(1 - 2t_{1}\right)^{\lfloor \frac{n}{2} \rfloor}}{2}\text{.}
\end{align}
Hence, for all $n \in \left\{0, 1, 2, \dots \right\}$,
\begin{equation}
    \mathbb{E}\left(M_{2}^{n}\right) = \frac{t_{2}\left(2\lfloor \frac{n}{2} \rfloor t_{1} + \left(1 - 2t_{1}\right)^{\lfloor \frac{n}{2} \rfloor} - 1\right)}{4}\text{.}
\end{equation}

\subsubsection{Pairs Consumed to Produce $M'$ Pairs}\label{DEJMPSAsymptoticApproximation}

Alternatively, given $\left(k, m'\right) \in \left\{0, 1, 2, \dots\right\}^{2}$, we might be interested in the number of pairs from our pool that will be consumed in the production of $m'$ output pairs generated via $k$ iterations of the DEJMPS protocol, $I_{k}^{m'}$.
The distribution of this variable can also be calculated using the results of the previous section because it can only take even values and, for all $\left(k, m\right) \in \left\{0, 1, 2, \dots\right\}^{2}$ and $n \in \left\{0, 2, 4, \dots \right\}$, we have
\begin{equation}
    \Pr\left(I_{k}^{m} = n\right) = \Pr\left(M_{k}^{n} \geq m\right) - \Pr\left(M_{k}^{n-1} \geq m\right)\text{,}
\end{equation}
where $\Pr\left(M_{k}^{-1} \geq m\right) = 0$.
Alternatively, we could use the method of Appendix~\ref{IterativeMethod}, which is not restricted to the DEJMPS protocol.
\par
Having shown how we might calculate the exact distribution of the number of initial pairs required to purify $M'$ output pairs, we also now give a description of the asymptotic behaviour of this distribution in the limit as $M'$ goes to infinity.
For all $k \in \left\{0, 1, 2, \dots\right\}$, as $M' \to \infty$, $\frac{I_{k}^{M}}{M'}$ converges in distribution to a normal distribution with mean $\mu_{k}$ and variance $\frac{\sigma_{k}^{2}}{M'}$, where $\mu$ is the mean number of states the DEJMPS protocol with $k$ iterations consumes to produce one output state and $\sigma^{2}$ is the variance in the number of states consumed to do so.
Hence, for all $k$ and large $M'$, $I_{k}^{M'}$ is approximately distributed as a normal random variable with mean $M'\mu_{k}$ and variance $M'\sigma_{k}^{2}$.
\par
For $k \in \mathbb{N}$, to apply the DEJMPS protocol $k$ times, we must produce two pairs by applying it $k-1$ times and, if we fail to apply the DEJMPS protocol to them successfully, produce two more pairs by applying the DEJMPS protocol $k-1$ times and so on until we succeed.
Letting $J_{k}$ denote the number of pairs we produce by applying the DEJMPS protocol $k-1$ times before ultimately succeeding, we have
\begin{align}
    \mu_{k} &= \sum_{n = 1}^{\infty}{P\left(J_{k} = 2n\right)\mathbb{E}\left(I_{k}^{1} \middle \vert J_{k}=2n\right)} \nonumber \\
    &= \sum_{n=1}^{\infty}{2t_{k}\left(1-t_{k}\right)^{n-1}n\mu_{k-1}} \nonumber \\
    &= \frac{2\mu_{k-1}}{t_{k}}\text{.}
\end{align}
Then, by induction, we have, for all $k \in \left\{0, 1, \dots\right\}$,
\begin{equation}
\mu_{k} = \frac{2^{k}}{s_{k}}\text{,}
\end{equation}
where $s_{0} = 1$ and, for all $k' \in \mathbb{N}$, $s_{k'} = t_{k'}s_{k'-1} = \prod_{i=1}^{k'}t_{i}$.
\par
Similarly, for all $k \in \left\{0, 1, 2 \dots \right\}$, we let $\sigma_{k}^{2} = \mathrm{Var}\left(I_{k}^{1}\right)$.
Then, by the law of total variance, for every $k \in \mathbb{N}$, we have
\begin{align}
    \sigma_{k}^{2} &= \mathbb{E}\left(\mathrm{Var}\left(I_{k}^{1} \middle \vert J_{k}\right)\right) + \mathrm{Var}\left(\mathbb{E}\left(I_{k}^{1} \middle \vert J_{k}\right)\right) \nonumber \\
    &= \frac{2\sigma_{k-1}^{2}}{t_{k}} + \frac{2^{2k}\mathrm{Var}\left(\frac{J_{k}}{2}\right)}{s_{k-1}^{2}} \nonumber \\
    &= \frac{2\sigma_{k-1}^{2}}{t_{k}} + \frac{2^{2k}\left(1-t_{k}\right)}{s_{k}^{2}}.
\end{align}
We can then see by induction that $\sigma_{0}^{2} = 0$ and, for all $k \in \mathbb{N}$, we have 
\begin{equation}
    \sigma_{k}^{2} = \frac{2^{k+1}\left(2^{k-1} - s_{k}\left(1+ \sum_{i=1}^{n-1}{\frac{2^{i-1}}{s_{i}}}\right)\right)}{s_{k}^{2}}\text{.}
\end{equation}

\section{Calculating $\Pr\left(X_{M'} = k, Y_{M'} = 1\right)$}\label{ExactCalculation}

\subsection{The Markov Chain Method} One way of calculating $\Pr\left(X_{M'}=k, Y_{M'}=1\right)$ for different values of $k$ and $M'$ is to characterise the process of repeatedly applying our protocol as a Markov chain in which each step corresponds to consuming one more state from our initial pool.
Then, we find the probability distribution for the Markov chain's state after $N$ steps, from which we can find the probability distribution for the number of purified pairs produced by our protocol and the protocol used to produce the final pair by the time it has consumed $N$ pairs from the pool.
Of course, the exact specifications of the Markov chain will depend on the protocol we are working with.
We will now aim to make this method clear by looking at the Markov chain in the case in which we apply one of our interpolated DEJMPS protocols, although it could be straightforwardly adapted to general interpolated entanglement purification recurrence protocols.
\par
Suppose that the two protocols we are interpolating between involve applying the DEJMPS protocol $i$ and $j$ times with $i < j$ and we apply the former protocol with probability $p$.
Now, let the maximum number of states we could produce in this way from our initial $N$ states be denoted by $T = \lfloor \frac{N}{2^{i}}\rfloor$.
In total, our Markov chain will then have $\left(2T-1\right)\left(2^{i}+2^{j}+2\right)+2$ states, as follows.
\par
Two states correspond to us having successfully produced $T$ pairs.
The first of these corresponds having produced the $T$th pair by performing $i$ iterations of the DEJMPS protocol and the second corresponds to having produced the $T$th pair by performing $j$ iterations of the DEJMPS protocol.
Both of these states are absorbing as there is no hope of us producing any more pairs using only $N$ initial pairs.
\par
Suppose we wish to apply $i$ iterations of the DEJMPS protocol.
Along the way, there are $2^{i-1}$ stages we must pass through that we must consume input pairs to move on from.
The first of these is the stage where we have produced nothing.
The second is the stage where we have produced one pair via one iteration of the DEJMPS protocol.
The third is the stage where we have produced one pair via two iterations of the DEJMPS protocol.
The fourth is the stage where we have produced one pair via two iterations of the DEJMPS protocol and one pair via one iteration of the DEJMPS protocol.
The fifth is the stage where we have produced one pair via three iterations of the DEJMPS protocol.
This continues until the $2^{i-1}$th stage where we have produced one pair via $i-1$ iterations of the DEJMPS protocol, one via $i-2$ iterations of the DEJMPS protocol and so on, down to one pair via $1$ iteration of the DEJMPS protocol.
\par
Now, the DEJMPS protocol consumes two pairs from our initial pool each time it moves between two of the above stages.
Therefore, to ensure that each transition in the Markov chain corresponds to using one more pair from the initial pool of states, we must, in fact, associate to each of the above stages not one but two states of the Markov chain, one corresponding to being at that stage and another which transitions into the former with probability $1$.
Associated with these stages, then, we have a total of $2^{i}$ states in our Markov chain.
\par
On top of this, for $n \in \left\{0, 1, 2, \dots, T-1\right\}$, we need states from which we transition to a state corresponding to having produced our $n+1$th purified pair.
In fact, we must have two such states for each $n$, one corresponding to us being about to produce the $n+1$th pair via $i$ iterations of the DEJMPS protocol and one corresponding to us being about to produce the $n+1$th pair via $j$ iterations of the DEJMPS protocol.
This gives us $2^{i}+2^{j}+2$ states corresponding to having produced $n$ purified pairs and being in the process of producing our $n+1$th.
However, to keep track of whether our most recent pair was produced via $i$ iterations or $j$ iterations of the DEJMPS protocol, we actually need twice as many states, unless $n = 0$.
\par
In total, then, we have, as claimed, $\left(2T-1\right)\left(2^{i}+2^{j}+2\right) + 2$ states in our Markov chain.
Each state corresponds to the two parties having produced a particular number of purified pairs, being at a certain stage in the production of the next pair and having produced their most recent state via a certain number of DEJMPS iterations.
Each transition corresponds to one more pair being consumed from the pool of initial states.
\par
The transition probabilities of the Markov chain are determined by the success probabilities of different iterations of the DEJMPS protocol.
The probability distribution for the initial state reflects the fact that we start by trying to produce a state via $i$ iterations of the DEJMPS protocol with probability $p_{i}$ and by trying to produce a state via $j$ iterations of the DEJMPs protocol with probability $p_{j} = 1 - p_{i}$.
Then, the probability distribution for the state of the Markov chain after $N$ steps allows us to calculate $\Pr\left(X_{M'} = k, Y_{M'} = 1\right)$ for different values of $k$ and $M'$.

\begin{figure*}[t]
\centering
\includegraphics[width=1\textwidth]{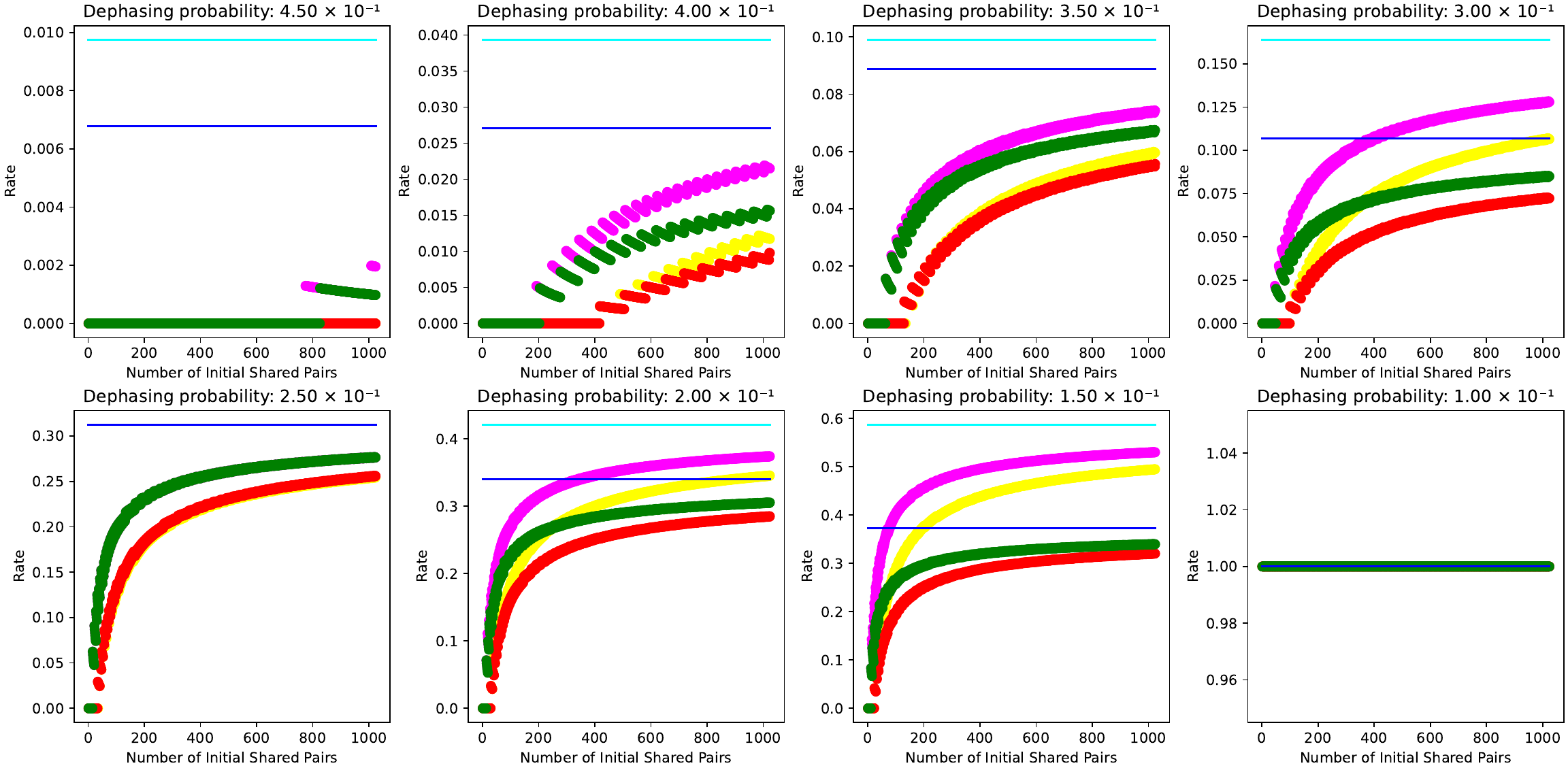}
\caption{The same as Figure~\ref{fig:FiniteSizeDepolarising}, except the initial states were generated by applying a dephasing channel to the chosen Bell state.}
\label{fig:FiniteSizeDephasing}
\end{figure*}

\begin{figure*}[t]
\centering
\includegraphics[width=1\textwidth]{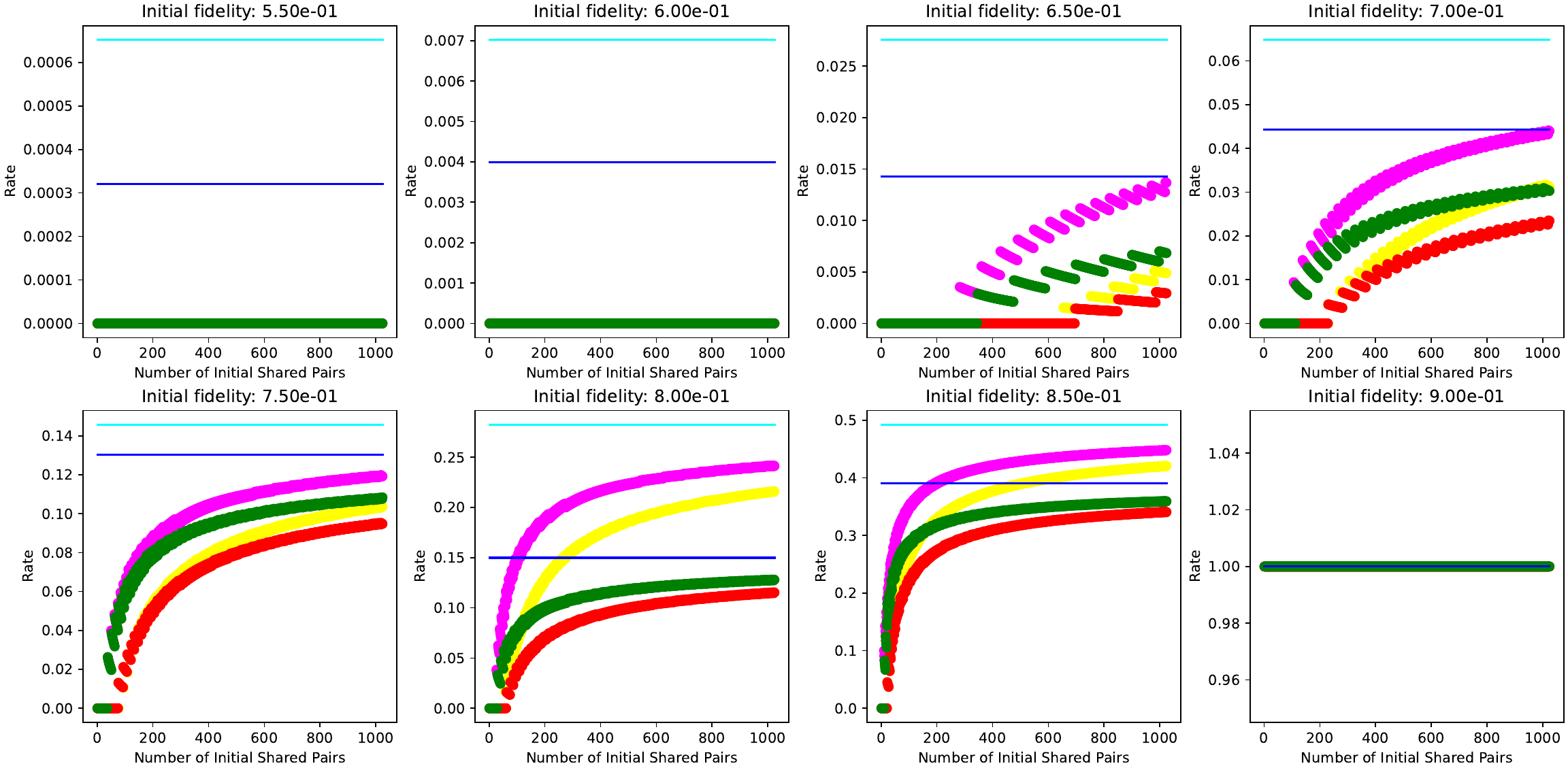}
\caption{The same as Figure~\ref{fig:FiniteSizeDepolarising}, except the initial states were generated by applying a Pauli channel where, conditional on the channel affecting the state, a phase flip is applied with probability $\frac{1}{2}$, a bit flip with probability $\frac{1}{3}$ and a bit and phase flip with probability $\frac{1}{6}$ to the chosen Bell state.}
\label{fig:FiniteSizePauli}
\end{figure*}

\begin{figure*}[t]
\centering
\includegraphics[width=1\textwidth]{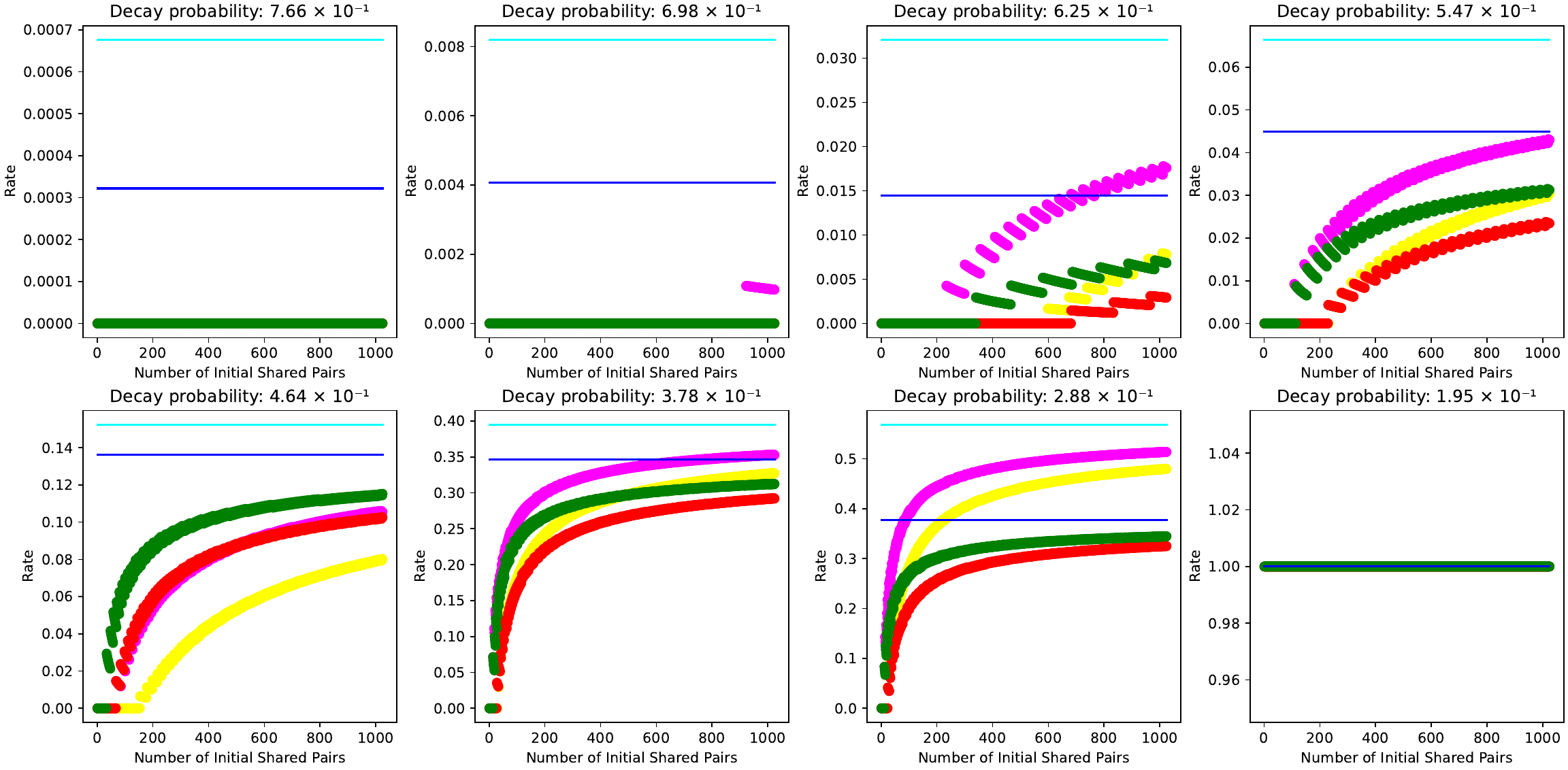}
\caption{The same as Figure~\ref{fig:FiniteSizeDepolarising}, except the initial states were generated by applying an amplitude damping channel to one half of the chosen Bell state.}
\label{fig:FiniteSizeAmplitudeDamping}
\end{figure*}

\subsection{The Iterative Method}\label{IterativeMethod}

We will now describe an alternative method which is sometimes faster than the Markov chain method described above.
Given $n \in \left\{0, 1, 2, \dots \right\}$, imagine we are trying to purify a single pair from a pool of $n$ initial pairs using our interpolated protocol and let $Y_{1, n} = 1$ if we succeed and $Y_{1, n} = 0$ otherwise.
Moreover, imagining that we have an infinite pool of initial states, for $M' \in \left\{0, 1, 2, \dots \right\}$, let $I^{M'}$ denote the number of states we consume in the process of distilling $M'$ pairs.
Then, for every $M' \in \mathbb{N}$ and $k \in \left\{1, 2, \dots, K\right\}$, we can calculate $\Pr\left(Y_{M'} = 1, X_{M'} = k\right)$ as 
\begin{equation}
    p_{k}\sum_{n=0}^{N}\Pr\left(I^{M'-1} = n\right)\Pr\left(Y_{1, N-n} = 1 \mid X_{1} = k\right)\text{.}
\end{equation}
\par
We might calculate $\Pr\left(Y_{1, N-n} = 1 \mid X_{1} = k\right)$ for different values of $k$ and $n$ using the Markov chain method described above.
Then, we note that, for every $M' \in \mathbb{N}$ and $n \in \left\{0, 1, 2, \dots \right\}$, we can  calculate $\Pr\left(I^{M'} = n\right)$ as
\begin{equation}
     \sum_{n' = 0}^{n-1}\Pr\left(I^{M'-1} = n'\right)\Pr\left(I^{1} = n - n'\right) 
\end{equation}
and, for every $n' \in \mathbb{N}$,
\begin{equation}
    \Pr\left(I^{1} = n'\right) = \Pr\left(Y_{1, n'} = 1\right) - \Pr\left(Y_{1, n' -1} = 1\right)\text{.}
\end{equation}
Therefore, we can iterate over the values of $M'$ we must consider and calculate the distribution of $I_{M'}$ for each of them, allowing us to find $M$.

\section{Additional Finite-Size Plots}\label{ExtraPlots}

Figures~\ref{fig:FiniteSizeDepolarising} and~\ref{fig:FiniteSizeDepolarisingPerPair} of the main text showed the performance of DEJMPS and interpolated DEJMPS protocols when our initial states are produced by applying a depolarizing channel to one half of a Bell pair.
As further examples, we show in Figures \ref{fig:FiniteSizeDephasing} to \ref{fig:FiniteSizeAmplitudeDamping} corresponding results for examples of other standard qubit channels, namely dephasing channels, Pauli channels and amplitude damping channels.
Note that in all of these cases, we assume that, before applying the very first iteration of the DEJMPS protocol, whether or not they are interpolating between different numbers of iterations, Alice and Bob apply local unitary rotations to permute their Bell states so as to maximize the Bell fidelity of the output states upon success of one iteration of the DEJMPS protocol.
This ensures our results are invariant under a relabelling of the Bell states.

\widetext

\clearpage
\newpage
\bibliography{references}

@article{QuantumTeleportation,
  title = {Teleporting an unknown quantum state via dual classical and Einstein-Podolsky-Rosen channels},
  author = {Bennett, Charles H. and Brassard, Gilles and Cr\'epeau, Claude and Jozsa, Richard and Peres, Asher and Wootters, William K.},
  journal = {Phys. Rev. Lett.},
  volume = {70},
  issue = {13},
  pages = {1895--1899},
  numpages = {0},
  year = {1993},
  month = {Mar},
  publisher = {American Physical Society},
  doi = {10.1103/PhysRevLett.70.1895},
  url = {https://link.aps.org/doi/10.1103/PhysRevLett.70.1895}
}

@article{SuperDenseCoding,
  title = {Communication via one- and two-particle operators on Einstein-Podolsky-Rosen states},
  author = {Bennett, Charles H. and Wiesner, Stephen J.},
  journal = {Phys. Rev. Lett.},
  volume = {69},
  issue = {20},
  pages = {2881--2884},
  numpages = {0},
  year = {1992},
  month = {Nov},
  publisher = {American Physical Society},
  doi = {10.1103/PhysRevLett.69.2881},
  url = {https://link.aps.org/doi/10.1103/PhysRevLett.69.2881}
}

@article{BB84,
   title={Quantum cryptography: Public key distribution and coin tossing},
   volume={560},
   ISSN={0304-3975},
   url={http://dx.doi.org/10.1016/j.tcs.2014.05.025},
   DOI={10.1016/j.tcs.2014.05.025},
   journal={Theoretical Computer Science},
   publisher={Elsevier BV},
   author={Bennett, Charles H. and Brassard, Gilles},
   year={2014},
   month=dec, pages={7–11} }

@article{EntanglementReview,
   title={Quantum entanglement},
   volume={81},
   ISSN={1539-0756},
   url={http://dx.doi.org/10.1103/RevModPhys.81.865},
   DOI={10.1103/revmodphys.81.865},
   number={2},
   journal={Reviews of Modern Physics},
   publisher={American Physical Society (APS)},
   author={Horodecki, Ryszard and Horodecki, Paweł and Horodecki, Michał and Horodecki, Karol},
   year={2009},
   month=jun, pages={865–942} }

@article{EntanglementOfFormation,
   title={Mixed-state entanglement and quantum error correction},
   volume={54},
   ISSN={1094-1622},
   url={http://dx.doi.org/10.1103/PhysRevA.54.3824},
   DOI={10.1103/physreva.54.3824},
   number={5},
   journal={Physical Review A},
   publisher={American Physical Society (APS)},
   author={Bennett, Charles H. and DiVincenzo, David P. and Smolin, John A. and Wootters, William K.},
   year={1996},
   month=nov, pages={3824–3851} }

@article{ContinuousDistribution,
   title={Performance metrics for the continuous distribution of entanglement in multiuser quantum networks},
   volume={108},
   ISSN={2469-9934},
   url={http://dx.doi.org/10.1103/PhysRevA.108.052615},
   DOI={10.1103/physreva.108.052615},
   number={5},
   journal={Physical Review A},
   publisher={American Physical Society (APS)},
   author={Iñesta, {\'A}lvaro G. and Wehner, Stephanie},
   year={2023},
   month=nov, pages={052615}}

@misc{RateBottleneck,
      title={Practical Routing and Criticality in Large-Scale Quantum Communication Networks}, 
      author={Cillian Harney and Stefano Pirandola},
      year={2025},
      eprint={2509.10908},
      archivePrefix={arXiv},
      primaryClass={quant-ph},
      url={https://arxiv.org/abs/2509.10908}, 
}

@article{DistillationInterpolation,
   title={Optimizing practical entanglement distillation},
   volume={97},
   ISSN={2469-9934},
   url={http://dx.doi.org/10.1103/PhysRevA.97.062333},
   DOI={10.1103/physreva.97.062333},
   number={6},
   journal={Physical Review A},
   publisher={American Physical Society (APS)},
   author={Rozpędek, Filip and Schiet, Thomas and Thinh, Le Phuc and Elkouss, David and Doherty, Andrew C. and Wehner, Stephanie},
   year={2018},
   month=jun, pages={062333} }

@article{DEJMPS,
   title={Quantum Privacy Amplification and the Security of Quantum Cryptography over Noisy Channels},
   volume={77},
   ISSN={1079-7114},
   url={http://dx.doi.org/10.1103/PhysRevLett.77.2818},
   DOI={10.1103/physrevlett.77.2818},
   number={13},
   journal={Physical Review Letters},
   publisher={American Physical Society (APS)},
   author={Deutsch, David and Ekert, Artur and Jozsa, Richard and Macchiavello, Chiara and Popescu, Sandu and Sanpera, Anna},
   year={1996},
   month=sep, pages={2818–2821} }

@article{WernerState,
  title = {Quantum states with Einstein-Podolsky-Rosen correlations admitting a hidden-variable model},
  author = {Werner, Reinhard F.},
  journal = {Phys. Rev. A},
  volume = {40},
  issue = {8},
  pages = {4277--4281},
  numpages = {0},
  year = {1989},
  month = {Oct},
  publisher = {American Physical Society},
  doi = {10.1103/PhysRevA.40.4277},
  url = {https://link.aps.org/doi/10.1103/PhysRevA.40.4277}
}

@article{BBPSSW,
   title={Purification of Noisy Entanglement and Faithful Teleportation via Noisy Channels},
   volume={76},
   ISSN={1079-7114},
   url={http://dx.doi.org/10.1103/PhysRevLett.76.722},
   DOI={10.1103/physrevlett.76.722},
   number={5},
   journal={Physical Review Letters},
   publisher={American Physical Society (APS)},
   author={Bennett, Charles H. and Brassard, Gilles and Popescu, Sandu and Schumacher, Benjamin and Smolin, John A. and Wootters, William K.},
   year={1996},
   month=jan, pages={722–725} }

@article{RelativeEntropyOfEntanglement,
  title = {Entanglement measures and purification procedures},
  author = {Vedral, V. and Plenio, M. B.},
  journal = {Phys. Rev. A},
  volume = {57},
  issue = {3},
  pages = {1619--1633},
  numpages = {0},
  year = {1998},
  month = {Mar},
  publisher = {American Physical Society},
  doi = {10.1103/PhysRevA.57.1619},
  url = {https://link.aps.org/doi/10.1103/PhysRevA.57.1619}
}

@article{AsymptoticRelativeEntropyOfEntanglement,
   title={Asymptotic Relative Entropy of Entanglement},
   volume={87},
   ISSN={1079-7114},
   url={http://dx.doi.org/10.1103/PhysRevLett.87.217902},
   DOI={10.1103/physrevlett.87.217902},
   number={21},
   journal={Physical Review Letters},
   publisher={American Physical Society (APS)},
   author={Audenaert, K. and Eisert, J. and Jané, E. and Plenio, M. B. and Virmani, S. and De Moor, B.},
   year={2001},
   month=nov, pages={217902}}

@article{PurificationReview,
   title={Entanglement purification and quantum error correction},
   volume={70},
   ISSN={1361-6633},
   url={http://dx.doi.org/10.1088/0034-4885/70/8/R03},
   DOI={10.1088/0034-4885/70/8/r03},
   number={8},
   journal={Reports on Progress in Physics},
   publisher={IOP Publishing},
   author={Dür, W and Briegel, H J},
   year={2007},
   month=jul, pages={1381–1424} }

@article{pirandola2017fundamental,
  title={Fundamental limits of repeaterless quantum communications},
  author={Pirandola, Stefano and Laurenza, Riccardo and Ottaviani, Carlo and Banchi, Leonardo},
  journal={Nature communications},
  volume={8},
  number={15043},
  year={2017},
  publisher={Nature Publishing Group}
}

@article{pirandola2020advances,
  title={Advances in quantum cryptography},
  author={Pirandola, Stefano and Andersen, Ulrik L and Banchi, Leonardo and Berta, Mario and Bunandar, Darius and Colbeck, Roger and Englund, Dirk and Gehring, Tobias and Lupo, Cosmo and Ottaviani, Carlo and others},
  journal={Advances in optics and photonics},
  volume={12},
  number={4},
  pages={1012--1236},
  year={2020},
  publisher={Optica Publishing Group}
}

@Article{Ekert1991,
  Title                    = {Quantum cryptography based on Bell's theorem},
  Author                   = {Ekert, Artur K},
  Journal                  = {Physical Review Letters},
  Year                     = {1991},
  Number                   = {6},
  Pages                    = {661},
  Volume                   = {67},
  Publisher                = {APS}
}

@Article{Pirandola2019,
  author    = {Pirandola, Stefano},
  journal   = {Communications Physics},
  title     = {End-to-end Capacities of a Quantum Communication Network},
  year      = {2019},
  pages     = {51},
  volume    = {2},
  doi       = {10.1038/s42005-019-0147-3},
  publisher = {Springer},
}

@article{Shchukin2019,
  title = {Waiting time in quantum repeaters with probabilistic entanglement swapping},
  author = {Shchukin, E. and Schmidt, F. and van Loock, P.},
  journal = {Phys. Rev. A},
  volume = {100},
  issue = {3},
  pages = {032322},
  numpages = {20},
  year = {2019},
  month = {Sep},
  publisher = {American Physical Society},
  doi = {10.1103/PhysRevA.100.032322},
  url = {https://link.aps.org/doi/10.1103/PhysRevA.100.032322}
}

@article{Lopiparo2020,
  title = {Aggregating quantum networks},
  author = {Lo~Piparo, Nicolo and Hanks, Michael and Nemoto, Kae and Munro, William J.},
  journal = {Phys. Rev. A},
  volume = {102},
  issue = {5},
  pages = {052613},
  numpages = {9},
  year = {2020},
  month = {Nov},
  publisher = {American Physical Society},
  doi = {10.1103/PhysRevA.102.052613},
  url = {https://link.aps.org/doi/10.1103/PhysRevA.102.052613}
}

@misc{Lopiparo2025,
      title={Quality of Service in aggregated quantum networks}, 
      author={Lo~Piparo, Nicolò  and J. Munro, William and Nemoto, Kae},
      year={2025},
      eprint={2501.18846},
      archivePrefix={arXiv},
      primaryClass={quant-ph},
      url={https://arxiv.org/abs/2501.18846}, 
}

@article{Harney2022,
  title = {Analytical Methods for High-Rate Global Quantum Networks},
  author = {Harney, Cillian and Pirandola, Stefano},
  journal = {PRX Quantum},
  volume = {3},
  issue = {1},
  pages = {010349},
  numpages = {12},
  year = {2022},
  month = {Mar},
  publisher = {American Physical Society},
  doi = {10.1103/PRXQuantum.3.010349},
  url = {https://link.aps.org/doi/10.1103/PRXQuantum.3.010349}
}

@article{Harney2022b,
  title = {End-To-End Capacities of Hybrid Quantum Networks},
  author = {Harney, Cillian and Fletcher, Alasdair I. and Pirandola, Stefano},
  journal = {Phys. Rev. Appl.},
  volume = {18},
  issue = {1},
  pages = {014012},
  numpages = {20},
  year = {2022},
  month = {Jul},
  publisher = {American Physical Society},
  doi = {10.1103/PhysRevApplied.18.014012},
  url = {https://link.aps.org/doi/10.1103/PhysRevApplied.18.014012}
}

\end{document}